\documentclass[10pt,pra,aps,article,oneside,twocolumn,superscriptaddress,showpacs]{revtex4-2}
\usepackage[english]{babel}
\usepackage[latin1]{inputenc}
\usepackage[T1]{fontenc}
\usepackage{lmodern}
\usepackage{ulem}
\usepackage{dcolumn}
\usepackage{subfigure}
\usepackage{footnote}
\usepackage[title]{appendix}                            
\usepackage{multirow}
\usepackage{amsmath,amsfonts,amssymb}
\usepackage{graphicx}                                   
\usepackage{latexsym}                                   
\usepackage{color}
\usepackage{xcolor}
\usepackage{upgreek}                                    
\usepackage{hyperref}                                   
\graphicspath{{./Figures/}}

\newcommand{\Eqn}[2]{\begin{equation}\label{eq:#1}#2\end{equation}}
\newcommand{\Aln}[2]{\begin{align}\label{eq:#1}#2\end{align}}
\newcommand{\RefEqg}[1]{\textcolor{magenta!50!black}{(\ref{eq:#1})}}
\newcommand{\Fig}[3]{\begin{figure}[t]\centerline{\includegraphics[width=#1 truecm]{Figures/#2}}\caption{\small{#3}}\label{fig:#2}\end{figure}}
\newcommand{\RefFig}[1]{\textcolor{magenta!50!black}{\ref{fig:#1}}}
\newcommand{\FndDoi}[1]{}
\newcommand{\FndPap}[1]{}

\newcommand{\U}[1]{\operatorname{#1}}					

\renewcommand{\Im}[0]{\mathfrak{Im\,}}

\newcommand{\Mtz}[1]{\left(\begin{matrix}#1\end{matrix}\right)}

\renewcommand{\>}{\rangle}
\renewcommand{\r}{\mathbf{r}}
\renewcommand{\k}{\mathbf{k}}

\renewcommand{\emph}[1]{\textit{#1}}

\begin{document}

\title{Photonic bands and normal mode splitting in optical lattices interacting with cavities}

\author{\firstname{Ph.W.} \surname{Courteille}}
\email{philippe.courteille@ifsc.usp.br}
\affiliation{Instituto de F\'isica de S\~ao Carlos,
Universidade de S\~ao Paulo, S\~ao Carlos, SP 13566-970, Brazil}
\affiliation{Departamento de F\'isica,
Universidade Federal de S\~ao Carlos, S\~ao Carlos, SP 13565-905, Brazil}

\author{\firstname{D.} \surname{Rivero}}
\affiliation{Instituto de F\'isica de S\~ao Carlos,
Universidade de S\~ao Paulo, S\~ao Carlos, SP 13566-970, Brazil}

\author{\firstname{G.H.} \surname{de Fran\c{c}a}}
\affiliation{Instituto de F\'isica de S\~ao Carlos,
Universidade de S\~ao Paulo, S\~ao Carlos, SP 13566-970, Brazil}

\author{\firstname{C.A.} \surname{Pessoa Jr}}
\affiliation{Instituto de F\'isica de S\~ao Carlos,
Universidade de S\~ao Paulo, S\~ao Carlos, SP 13566-970, Brazil}

\author{\firstname{A.} \surname{Cipris}}
\affiliation{Instituto de F\'isica de S\~ao Carlos,
Universidade de S\~ao Paulo, S\~ao Carlos, SP 13566-970, Brazil}

\author{\firstname{M.} \surname{N\'u\~nez Portela}}
\affiliation{Laboratorio de \'Optica Cu\'antica, 
Universidad de los Andes, A.A. 4976, Bogot\'a, D.C., Colombia}

\author{\firstname{R.C.} \surname{Teixeira}}
\affiliation{Departamento de F\'isica,
Universidade Federal de S\~ao Carlos, S\~ao Carlos, SP 13565-905, Brazil}

\author{\firstname{S.} \surname{Slama}}
\affiliation{Center for Quantum Sciences and Physikalisches Institut, Eberhard-Karls Universit\"at T\"ubingen, 72076 T\"ubingen, Germany}

\begin{abstract}
Strong collective interaction of atoms with an optical cavity causes normal mode splitting of the cavity's resonances, whose width is given by the collective coupling strength. At low optical density of the atomic cloud the intensity distribution of light in the cavity is ruled by the cavity's mode function, which is solely determined by its geometry. In this regime the dynamics of the coupled atom-cavity system is conveniently described by the open Dicke model, which we apply to calculating normal mode splitting generated by periodically ordered clouds in linear and ring cavities. We also show how to use normal mode splitting as witness for Wannier-Bloch oscillations in the tight-binding limit. At high optical density the atomic distribution contributes to shaping the mode function. This regime escapes the open Dicke model, but can be treated by a transfer matrix model provided the saturation parameter is low. Applying this latter model to an atomic cloud periodically ordered into a one-dimensional lattice, we observe the formation of photonic bands gaps competing with the normal mode splitting. We discuss the limitations of both models and point out possible pathways to generalized theories.
\end{abstract}

\pacs{
     xxx,  
}
\maketitle
\tableofcontents

\section{Introduction} \label{sec:Intro}

Popular models concerned with the interaction of large atomic ensembles with light modes are the Open Dicke model (ODM) \cite{Dicke54,Garraway11,Kirton19}, the Coupled Dipoles Model (CDM) \cite{Friedberg73,Scully06,Courteille10}, and the Transfer Matrix Model (TMM) \cite{Born-80}. Every model focuses on a different aspect of the coupled system and to this end applies a different simplifying assumption. Its applicability therefore depends on the regime in which the coupling is investigated, for example, weak or strong collective coupling, small or large saturation, and low or high optical density. Each approach has its limitations, advantages, and disadvantages, and its preference depends on the focus of the investigation.

The ODM has been highly useful for unraveling how atomic ensembles collectively interact with single light modes. The prediction of super- and subradiance are prominent examples \cite{Dicke54,Lehmberg70b,Gross82}. The basic idea underlying the model is that the atoms are indistinguishable with respect to their interaction with the light mode, so that excitation of an individual atomic spin can be described as a step up the ladder formed by the eigenstates of a collective spin. This trick permits a dramatic reduction of the dimension of the collective Hilbert space, but the price to pay is a loss of individual addressability of the atoms. Furthermore, the ODM only applies to situations in which the coupling strength between the light mode and an atom solely depends on its location within the mode volume, but not the location or dynamics of the other atoms. This precludes the applicability of the ODM to optically dense clouds. Nevertheless, the dynamics of atomic clouds interacting with the light modes supported by \emph{optical cavities} is usually described by the ODM \cite{Norcia18b,LiZ22,Rivero23}.

On the other hand, the TMM is a linear model describing one-dimensional propagation of light through consecutive layers of scatterers or optical elements, which can be optically dilute or dense, and has been successfully applied to describe Bragg reflection and the formation of forbidden photonic bands in one-dimensional optical lattices generated by two counter-propagating laser beams in free space \cite{Deutsch95b,Slama06,Schilke11}. It is a characteristic of optical cavities to enforce a one-dimensional geometry, which is for many systems a sufficient approximation. Furthermore, laser-pumped cavities can sustain optical lattices. It is thus an interesting question, to what extend the TMM can be applied to interacting atom-cavity systems and identify circumstances in which it even reaches beyond the ODM, especially in cases where atomic (dis-) ordering is expected to have an impact on the coupled dynamics or when the atomic cloud is optically dense.

Finally, the CDM has been fruitfully applied to phenomena in the limit of weak excitation, particularly to situations where the arrangement of atoms in space (ordered or disordered) plays an important role \cite{Svidzinsky10,Bienaime11}. While the CDM facilitated studies of single-photon super- and subradiance in large clouds of atoms \cite{Araujo16,Guerin16}, it is difficult to accommodate with the presence of surfaces or even cavities \cite{JonesR18,Araujo24}, and we will not use it in this work.

\bigskip

In this work, considering specific experimental situations, we compare the ODM and the TMM when applied to atoms trapped within the optical mode of a linear cavity and a ring cavity in order to illustrate their capabilities and limitations. We report three main results: (1)~Applying the ODM to periodically ordered atoms in a linear and in a ring cavity, we calculate the dependence of the \textit{normal mode splitting} on the lattice constant. We find that, in both cases, the dependence is fully characterized by a single complex parameter, which is the bunching parameter. Hence, measuring the bunching parameter via observation of the normal mode splitting yields information on the periodic ordering, which can be exploited to monitor variations of the periodic ordering, e.g.~due to Bloch oscillations \cite{ZhangH23}.

(2)~We apply the TMM to periodically ordered atoms in a linear cavity and extend the model to ring cavities. We benchmark the model with the ODM in the limit of low optical density of the atomic cloud, but we also show that validity of the TMM reaches out into the regime of high optical density, inaccessible to the ODM. Indeed, while the ODM presupposes that the atoms only interact via their coupling to the same mode function of the cavity \cite{Bychek21}, in dense clouds the atoms can interact by exchanging photons directly, thus bypassing the cavity mode. A dense and disordered atomic cloud will, due to absorption, shape the intensity profile along the cavity's optical axis producing a shadow on atoms located further downstream the light beam's energy flux. If a cloud is dense and periodically ordered, we expect the formation of \textit{photonic stop bands}, i.e.~frequency bands inside which the propagation of light is prevented \cite{Deutsch95b,Coevorden96,Slama06,Antezza09,YuDeshui11,Schilke11,Schilke12,Samoylova14,Samoylova14b}. Inside the optical lattice the light intensity can be considerably enhanced due to multiple Bragg reflections, while behind the lattice it is attenuated. Thus, in the optically dense regime the atomic cloud participates in shaping the mode structure with dramatic impact on the normal mode spectra of the coupled atom-cavity system. While these features are grasped by the TMM, it is nonetheless important to point out that the TMM does not allow incorporation of effects due to saturation nor feedback.

(3)~The coupled atom-cavity system intertwines three manifestation of photonic stop bands: The cavity spectrum itself described by an Airy function, the normal mode splitting resulting from cavity-atom interaction, and finally the photonic bandgap (PBG) resulting from multiple paths interference of the light propagating inside the optical lattice. Paths covering different distances inside the lattice correspond to different round-trip times in the cavity. We will discuss how the cavity can be employed to filter particular paths out of the manifold.

At the end, we will discuss the limitations of the presented models and suggest possible extensions toward more general theories.

\bigskip

The paper is organized as follows. In Sec.\,\ref{sec:Bunching}, based on the ODM model, we present simulations for optical lattices suspended in linear and ring cavities focusing on parameter regimes where the TMM yields \emph{identical} results. Sec.\,\ref{sec:BunchingWeak} introduces the model. In Sec.\,\ref{sec:BunchingNormal} we show simulations of normal mode splitting as a function of atomic bunching and de-bunching caused by incommensurate lattices and thermal spreading and compare them with previous results \cite{Rivero23}. Sec.\,\ref{sec:BunchingBlochoscis} proposes a possible application for the detection of Wannier-Bloch oscillations in the tight-binding regime \cite{ZhangH23}. In particular, we will show that the spreading of atomic wavepackets over several lattice sites can be probed via Bragg reflection.
	
The second part of the paper is Sec.\,\ref{sec:Transfer}, which focuses on the TMM for optical lattices in linear and ring cavities, showing simulations in parameter regimes where the TMM yields results \emph{differing} from those obtained with the ODM. In Sec.\,\ref{sec:TransferPolarizability} we work out the link between the models, and in Secs.\,\ref{sec:TransferTransfer} and \ref{sec:TransferRing} we expose, respectively, the linear and ring cavity transfer matrix formalism. 

In the third part, we discuss in Sec.\,\ref{sec:BandgapPropagation} the propagation of light through a cavity filled with a dense cloud and in Sec.\,\ref{sec:BandgapPhotonic} the interaction of PBGs with cavities, showing that the cavity filters out specific beam paths traversing the optical lattice thus allowing for a spectral analysis of the PBG. The paper concludes in Sec.\,\ref{sec:Conclusion} with a comparison of the ODM and the TMM and briefly points out possible pathways toward a complete quantum model holding for the saturated dense cloud regime.

\section{Open Dicke model} \label{sec:Bunching}

Ring cavities are fundamentally different from linear cavities in many ways. First of all, they support two energetically degenerate counter-propagating modes that share the same mode volume. As long as the modes are not coupled, their photon budgets remain independent. Furthermore, photon back-scattering processes conserve momentum, so that the photon number on each degenerate mode is coupled to the momentum state of an atom trapped within the cavity mode volume. If pumped by laser light in only one direction, the light intensity along the ring cavity's optical axis is almost constant over long distances. Hence, a laser beam far-detuned from an atomic resonance generates a one-dimensional constant dipolar potential capable of trapping a uniform atomic cloud. On the other hand, with light of wavelength $\lambda_\text{lat}$ injected from both sides into the cavity, a standing light wave with periodicity $\lambda_\text{lat}/2$ is formed, which can confine a cold atomic cloud with nearly perfect periodic ordering in a 1D optical lattice aligned with the cavity's optical axis. Figs.\,\RefFig{Scheme1}(a,b) show possible geometries for linear and ring cavities, and Fig.\,\RefFig{Scheme1}(c) illustrates the atomic density distribution over the optical lattice at finite temperature.
 	\Fig{8.7}{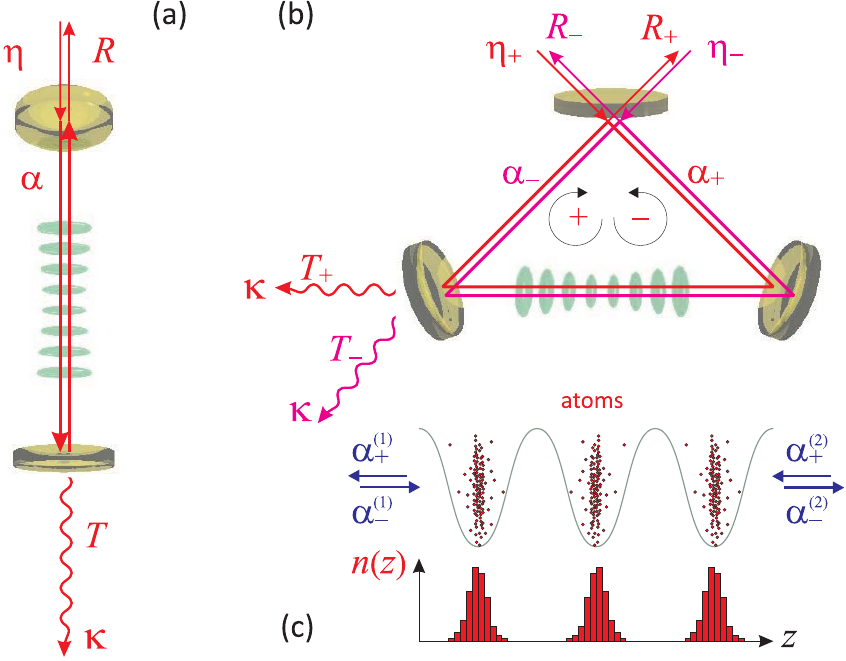}{(a)~Scheme of a linear cavity laser-pumped at a rate $\eta$ and afflicted by losses occurring at a rate $\kappa$, and
		containing an optical lattice of cold atoms.
		(b)~Scheme of a ring cavity with two counter-propagating modes rotating in clockwise (+) and counter-clockwise direction (-).
		(c)~Illustration of the atomic distribution in three adjacent sites of an optical lattice.}

\subsection{Weak excitation and the role of atomic bunching in linear and ring cavities} \label{sec:BunchingWeak}

The strength of the collective interaction between an atomic cloud and a cavity not only depends on the number $N$ of atoms but also on their individual coupling to the cavity's mode function through their location $z_j$ along the cavity axis. The mode function of linear cavities is sinusoidally modulated with a periodicity determined by the wavelength of a probe laser irradiated in resonance with a cavity mode and not far away from an atomic transition, $\lambda=2\pi/k$. For ring cavities the mode function has translationally invariant amplitude. Note that the mode volume can also depend on the electronic state of the atoms \cite{Suarez23}. We now define atomic \emph{bunching parameters},\Eqn
	{Bunching01}{b_0 \equiv \frac{1}{N}\sum_{j=1}^N\cos^2kz_j ~~~,~~~ b_\pm \equiv \frac{1}{N}\sum_{j=1}^Ne^{\pm 2\imath kz_j}}
for the linear and for the ring cavity, respectively. These parameters measure the longitudinal overlap of the periodically ordered atomic cloud with the cavity mode function at the resonant wavelength $\lambda$. For example, for a cloud homogeneously distributed along the cavity axis we get $b_0=\tfrac{1}{2}$ and $b_\pm=0$. In contrast, perfect bunching means that all atoms have the same distance modulo the lattice period $2(z_i-z_j)/\lambda_\text{lat}\in\mathbb{Z}$. In such cases, we find $b_0=\cos^2kz_j=\cos^2kz_0$, respectively, $b_\pm=e^{2\imath kz_j}=e^{2\imath kz_0}$, where $z_0$ represents the distance between the position of any atom and the nearest point of zero spatial phase of the light field.

In addition to the positions of the atoms $z_j$, the degrees of freedom of the coupled atom-cavity system include the amplitudes of the intracavity light, which are treated as classical fields and denoted as $\alpha$ for the linear cavity and $\alpha_\pm$ for the two counter-propagating modes of a ring cavity. They are normalized to the electric field $\mathcal{E}_1$ generated by a single photon in the cavity mode, so that $|\alpha|^2$ denotes the average number of photons. We do not consider atomic motion nor photonic recoil, assuming the optical lattice potential to be so deep and the binding so tight, that the atomic motion is not affected by the probe light. The equations of motion describing the coupled system can be derived in various ways. One approach is to formulate the collective Dicke Hamiltonian for $N$ atoms interacting with one or two counter-propagating cavity modes [see for example Eq.\,(11) in Ref.\,\cite{Rivero23}], identify the relevant dissipation mechanisms, and solve the master equation of the so-called open Dicke model (ODM) \cite{Dicke54,Garraway11,Kirton19}. By neglecting any type of quantum correlation and seeking stationary solutions, we find the following expressions, which allow us to calculate the intracavity light fields for given atomic positions in linear cavities,\Eqn
	{Bunching02}{\sum_j\frac{-U_\gamma\alpha\cos^2kz_j}{1+2|U_\gamma/g|^2|\alpha|^2\cos^2kz_j} = \imath\eta-\Delta_\kappa\alpha~,}
and for ring cavities (expression (22) of Ref.\,\cite{Rivero23}),\Eqn
	{Bunching03}{\sum_j\frac{-U_\gamma(\alpha_\pm+e^{\mp 2\imath kz_j}\alpha_\mp)}{1+2|U_\gamma/g|^2|e^{\imath kz_j}\alpha_++e^{-\imath kz_j}\alpha_-|^2}
		= \imath\eta_\pm-\Delta_\kappa\alpha_\pm~,}
where we defined the abbreviations,\Eqn
	{Bunching04}{U_\gamma \equiv U_0-\imath\gamma_0 \equiv \frac{g^2}{\Delta_\text{a}+\imath\Gamma/2} ~~~,~~~ \Delta_\kappa = \Delta_c+\imath\kappa~.}
Here, $\Gamma$ the decay width of the atomic transition, $\Delta_\text{a}$ the laser detuning from the atomic transition, $\kappa$ the cavity's field amplitude decay width, $\Delta_\text{c}$ the laser detuning from the nearest mode of the empty cavity, $U_0$ and $\gamma_0$ real parameters proportional to the real, respectively, imaginary part of the atomic linear electrical susceptibility, and $g$ is \emph{atom-field coupling strength} (equal to half the one-photon Rabi frequency),\Eqn
    {Bunching05}{2g = \frac{1}{\hbar}d\mathcal{E}_1
        = \sqrt{\frac{3\pi\Gamma\omega}{2k^3V_\text{m}}}
        = \sqrt{\frac{6\Gamma\delta_\text{fsr}}{k^2w^2}}~,}
where $d$ is the electric dipole moment of the atomic transition, $V_\text{m}=\tfrac{\pi}{2}Lw^2$ is the cavity mode volume, $L$ the cavity length, $w$ the Gaussian beam waist, and $\delta_\text{fsr}$ the free spectral range given in Hertz ($\delta_\text{fsr}=c/2L$ for a linear cavity and $\delta_\text{fsr}=c/L$ for a ring cavity). Finally, $\eta$ respectively $\eta_\pm$ are cavity pump rates proportional to the amplitudes of incident laser light, $\eta=\alpha_\text{in}\sqrt{\kappa\delta_\text{fsr}}$. Apart from $g$, a second important parameter ruling the atom-light interaction is the \emph{single-atom cooperativity},\Eqn
	{Bunching06}{\Upsilon = \frac{4g^2}{\kappa\Gamma} = \frac{F}{\pi}\frac{6}{k^2w^2}~,}
where $F\equiv\pi\delta_\text{fsr}/\kappa$ is called the finesse of the cavity.

The equations \RefEqg{Bunching02} and \RefEqg{Bunching03} are non-linear in the field amplitudes $\alpha$ and $\alpha_\pm$, respectively, and can only be solved analytically for particular cases, such as perfect atomic bunching or totally homogeneous clouds \cite{Rivero23}. Another analytically accessible case is low saturation, that is $|U_\gamma\alpha_\pm/g|\ll 1$. Then the denominators of the formulas \RefEqg{Bunching02} and \RefEqg{Bunching03} become equal to 1, and for the linear cavity we get immediately the solution,\Eqn
	{Bunching07}{\boxed{\alpha = \frac{\imath\eta}{\Delta_\kappa-NU_\gamma b_0}}}
while for a ring cavity,\Eqn
	{Bunching08}{\boxed{\alpha_\pm = \frac{\imath\eta_\pm(\Delta_\kappa-NU_\gamma)
		+\imath\eta_\mp NU_\gamma b_\mp}{(\Delta_\kappa-NU_\gamma)^2-N^2U_\gamma^2|b_+|^2}}~.}
We will soon see, how the imminent role of atomic bunching in the expressions \RefEqg{Bunching07} and \RefEqg{Bunching08} determines the shapes of transmission, reflection, and absorption spectra of the cavities. They are calculated from the amplitudes of the fields,\Eqn
	{Bunching09}{T = \left|\tfrac{\kappa\alpha}{\eta}\right|^2 ~~~,~~~ 
		R = \left|1-\tfrac{\kappa\alpha}{\eta}\right|^2 ~~~,~~~ 
		A = 1-T-R}
for the linear cavity and\Aln 
	{Bunching10}{T_\pm & = \left|\tfrac{\kappa\alpha_\pm}{\eta_+}\right|^2 ~~~,~~~ 
		R_\pm = \left|\tfrac{\eta_\pm}{\eta_+}-\tfrac{\kappa\alpha_\pm}{\eta_+}\right|^2 ~~~,\\
		A & = \sum_\pm\left(\tfrac{\eta_\pm}{\eta_+}-T_\pm-R_\pm\right)\nonumber}
for the ring cavity. We choose to normalize every transmission and reflection with the intensity $|\eta_+|^2$ of the light pumped into the cavity mode $\alpha_+$, because the examples to be discussed either assume $\eta_-=0$ or $\eta_-=\eta_+$. The transmission and reflection spectra are measured at the output ports indicated in Figs.\,\RefFig{Scheme1}(a,b).

\subsubsection{Parameter regimes for the simulations} \label{sec:BunchingNormalParameters}

For the sake of specificity, throughout the paper we will consider parameters close to those realized in our own experimental apparatus \cite{Rivero22,Rivero23}, which is dedicated to studies of the interaction between ultracold strontium clouds and a ring cavity. The probe laser (frequency $\omega$) is tuned close to the $^1S_0$-$^3P_1$ intercombination line at $\lambda_\text{a}=689\U{nm}$ in strontium atoms, whose transition linewidth is $\Gamma=(2\pi)\,7.4\U{kHz}$. The ring cavity is characterized by a mode beam diameter of $w\approx 70\U{\upmu m}$, a finesse of $F=1500$, a free spectral range on the order of $\delta_\text{fsr}\simeq 10^6\Gamma/2\pi$, an amplitude decay rate of $\kappa=(2\pi)\,3.4\U{MHz}$, and an atom-cavity coupling strength of roughly $g\simeq\Gamma$. Typically, $N\approx 200000$ atoms are stored in the standing light wave potential formed inside the ring cavity when it is pumped, in both directions, with laser light tuned to a cavity mode which can be very far away from the probe laser mode. The atoms then organize into a one-dimensional optical lattice with a periodicity given by half the resonant wavelength $\lambda_\text{lat}/2$, where they are then distributed over some $N_\text{s}\approx 300$ anti\-nodes. While most parameters will be kept fixed throughout the paper, others will be varied, in particular the coupling strength $g$, the finesse $F$ of the cavity, its configuration (linear or ring cavity), as well as the detunings of the probe laser from the atomic resonance $\Delta_\text{a}=\omega-\omega_\text{a}$, from the nearest cavity mode $\Delta_\text{c}=\omega-\omega_\text{c}$, and from the lattice laser $\Delta_\text{lat}=\omega-2\pi c/\lambda_\text{lat}$. Obviously, all results can be generalized to other atoms and arbitrary cavities.

By the fact that $\kappa\gg\Gamma$ our ring cavity operates deep in the so-called bad cavity limit. From a technical point of view, an interesting advantage of narrow atomic transitions is that a light frequency which is sufficiently detuned from atomic resonance to avoid spontaneous emission, may still be within the cavity's free spectral range. Hence, we can conveniently not only create a conservative dipolar light wave potential on adjacent cavity modes, but a \emph{standing} light wave potential (i.e.~an optical lattice) whose periodicity is nearly commensurate with the wavelength of resonant probe light. This is interesting, e.g.~for the creation and study of photonic band gaps \cite{Deutsch95b,Schilke11,Samoylova14b}.

\subsubsection{Debunching caused by incommensurate optical lattice} \label{sec:BunchingNormalDebunching}

The transmission and reflection spectra obtained from Eqs.\,\RefEqg{Bunching07} and \RefEqg{Bunching08}, respectively, critically depend on the degree of atomic bunching. Debunching of the atomic cloud can be caused by thermal motion (we will analyze this in the next section), but it can also be caused by an optical lattice whose periodicity is incommensurate with the wavelength of the probe light, $\lambda\ne\lambda_\text{lat}$. For a linear cavity, assuming that the atoms are with zero temperature (i.e.~located at the bottoms of the standing wave potential) and equally distributed over $N_\text{s}$ lattice sites, we describe this type of debunching by setting $z_j=j\lambda_\text{lat}/2+z_0$ in Eq.\,\RefEqg{Bunching01},\Aln
	{Bunching11}{b_0 & = \frac{1}{N_\text{s}}\sum_{j=(1-N_\text{s})/2}^{(N_\text{s}-1)/2}\cos^2(jk\lambda_\text{lat}/2+kz_0)\\
		& = \frac{1}{2}-\frac{\cos 2kz_0}{2N_\text{s}}\frac{\sin\frac{N_\text{s}}{2}k\lambda_\text{lat}}{\sin\frac{1}{2}k\lambda_\text{lat}}~,\nonumber}
and for a ring cavity by,\Aln
	{Bunching12}{b_\pm & = \frac{1}{N_\text{s}}\sum_{j=(1-N_\text{s})/2}^{(N_\text{s}-1)/2}e^{\pm 2\imath(jk\lambda_\text{lat}/2+kz_0)}\\
		& = \frac{e^{\pm 2\imath kz_0}}{N_\text{s}}\frac{\sin\frac{N_\text{s}}{2}k\lambda_\text{lat}}{\sin\frac{1}{2}k\lambda_\text{lat}}~.\nonumber}
The additional factor $kz_0$ allows to shift the overall phase of the optical lattice. The dependencies of the bunching parameters on the lattice detuning $\Delta_\text{lat}$ are shown in Fig.\,\RefFig{UncommensurateBunching}(a). Note that for small detunings,\Eqn
	{Bunching12b}{\frac{\sin\frac{N_\text{s}}{2}k\lambda_\text{lat}}{N_\text{s}\sin\frac{1}{2}k\lambda_\text{lat}} 
		\simeq \text{sinc\,}N_\text{s}(\pi-\tfrac{k\lambda_\text{lat}}{2})~.}
  	\Fig{8.7}{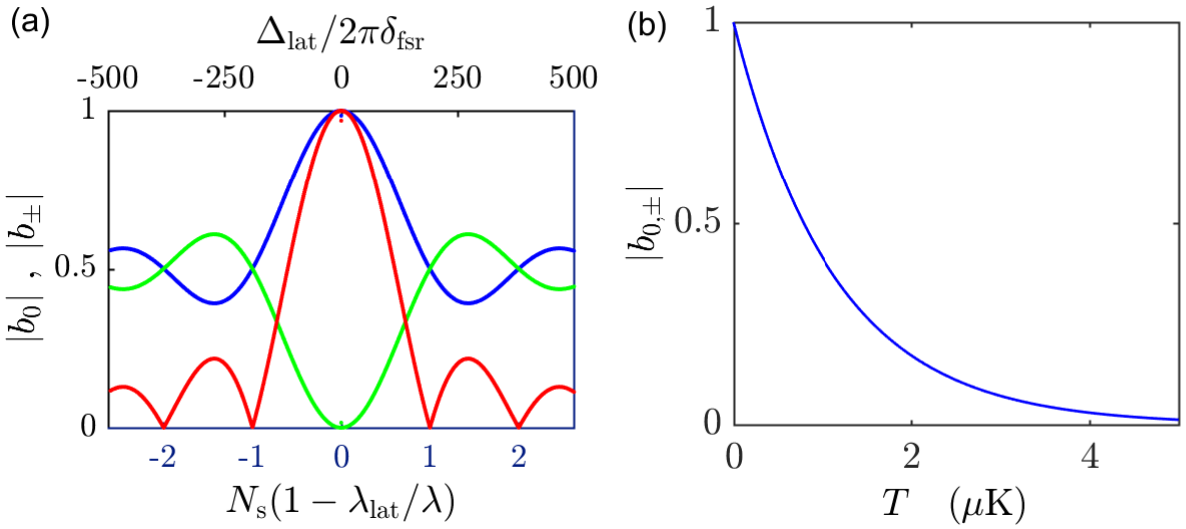}{(a)~Dependence of the bunching parameters on the lattice wavelength (bottom axis) and
		on the lattice detuning scaled to the free spectral range (top axis). The blue curve holds for a linear cavity 
		with $kz_0=0$ and the green curve with $kz_0=\pi/2$. The red curve holds for a ring cavity. The parameters are specified in 
		Sec.\,\ref{sec:BunchingNormalParameters}. (b)~Dependence of the bunching parameter for a ring cavity on temperature for a 
		lattice detuning fixed to $\Delta_\text{lat}=0$. Parameters as specified in Sec.\,\ref{sec:BunchingNormalTemperature}.}

\subsubsection{Debunching caused by finite temperature} \label{sec:BunchingNormalTemperature}

Until now we described the atomic layers of the optical lattice as perfectly thin infinite planes with transversally uniform density. However, when the atomic cloud is at finite temperature, the layers in each lattice site will have a finite extension along the optical axis [see Fig.\,\RefFig{Scheme1}(c)] whose width is given by the temperature of the cloud,\Eqn
	{Bunching13}{\bar z = \frac{1}{k}\sqrt{\frac{2k_BT_\text{at}}{V_0}}~,}
where $V_0$ is the depth of the optical lattice potential.

To calculate the temperature-induced reduction of the bunching parameter for an optical lattice extending over $N_\text{s}$ sites we assume a smooth longitudinal Gaussian density distribution modeled by,\Eqn
	{Bunching14}{n(z) = \frac{N}{N_\text{s}\bar z\sqrt{2\pi}}\sum_{j=(1-N_\text{s})/2}^{(N_\text{s}-1)/2}e^{-(z-j\lambda_\text{lat}/2)^2/2\bar z^2}~.}
and normalized to the total atom number $N$. For a linear cavity the bunching parameter is now given by the overlap integral,\Aln
	{Bunching15}{b_0 & = \frac{1}{N}\int_{-\infty}^\infty n(z)\cos^2(kz+kz_0)dz\\
		& = \frac{1}{2}-\frac{\cos 2kz_0}{2N_\text{s}}\frac{\sin\frac{N_\text{s}}{2}k\lambda_\text{lat}}
			{\sin\frac{1}{2}k\lambda_\text{lat}}e^{-2k^2\bar z^2}~.\nonumber}
Similarly, for a ring cavity the bunching parameter is given by the structure factor,\Aln
	{Bunching16}{b_\pm & = \frac{1}{N}\int_{-\infty}^\infty n(z)e^{\pm 2\imath(kz+kz_0)}dz\\
		& = \frac{e^{\pm 2\imath kz_0}}{N_\text{s}}\frac{\sin\frac{N_\text{s}}{2}k\lambda_\text{lat}}
			{\sin\frac{1}{2}k\lambda_\text{lat}}e^{-2k^2\bar z^2}~.\nonumber}
The Gaussian pre-factor in the expressions \RefEqg{Bunching15} and \RefEqg{Bunching16} is known as the Debye-Waller factor in crystallography. 

Fig.\,\RefFig{UncommensurateBunching}(b) shows the decrease of the bunching parameter with rising temperature for the case $\Delta_\text{lat}=0$. Assuming a temperature of $T_\text{at}=1\U{\upmu K}$ and $V_0=h\times 100\U{kHz}$, which are typical values, we find $k\bar z\approx 0.2$. This spread in position degrades the periodicity of the optical lattice with potentially important impact on absorption and phase shifts. Nevertheless, for the sake of clarity we will assume negligible thermal disorder, i.e.~$T_\text{at}=0$, for all calculations presented in the following except for those presented in Sec.\,\ref{sec:BandgapPropagationThermal}.

\subsection{Simulations of normal mode splittings} \label{sec:BunchingNormal} 

Fig.\,\RefFig{Linear_Bunching} shows transmission, reflection, and absorption profiles (obtained by scanning $\Delta_\text{c}$) for various detunings $\Delta_\text{lat}$ of the lattice wavelength calculated for a linear cavity from Eq.\,\RefEqg{Bunching07}. The normal modes appear as two distinct ridges with variable distance. The amount of normal mode splitting clearly depends on atomic bunching. For Figs.\,\RefFig{Linear_Bunching}(a-c) the atoms are localized at antinodes of the cavity mode when $\Delta_\text{lat}=0$ and for Figs.\,\RefFig{Linear_Bunching}(d-f) at nodes.
 	\Fig{8.7}{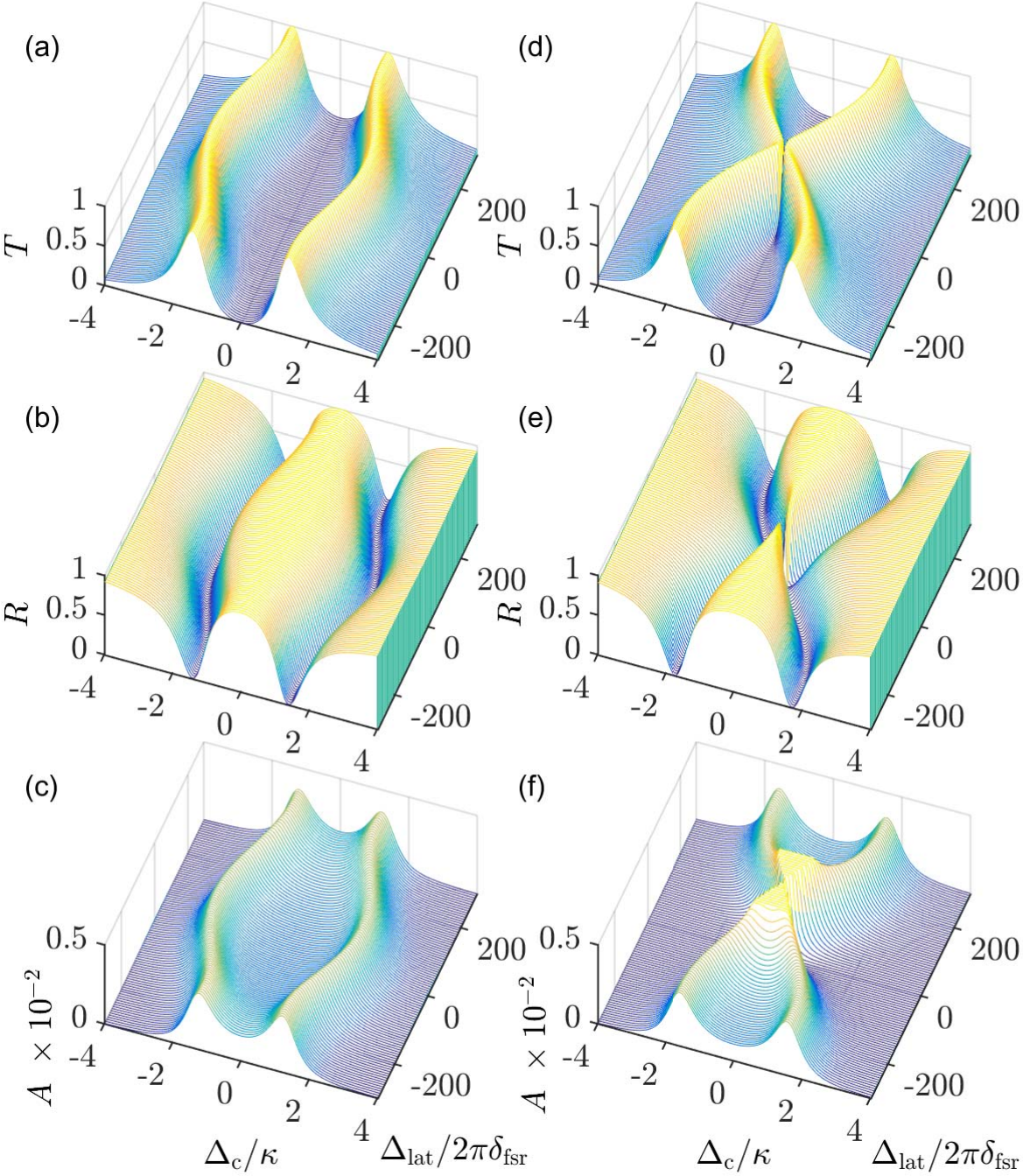}
		{Transmission $T$, reflection $R$, and absorption profiles $A=1-T-R$ for various detunings of the lattice wavelength for a linear cavity. 
		For the curves (a-c) at $\Delta_\text{lat}=0$ all atoms are located at an antinode of the optical lattice, $kz_j=0$, 
		while for the curves (d-f) they are located at a node, $kz_j=\pi/2$. All curves are calculated from Eq.\,\RefEqg{Bunching07} 
		using the ODM, but we note that using Eq.\,\RefEqg{Bandgap31} derived from the TMM we obtain exactly the same curves for the chosen 
		parameters: $\Delta_\text{ca}=0$, $N=5\cdot 10^5$, $T_\text{at}=0$. All other parameters as specified in Sec.\,\ref{sec:BunchingNormalParameters}.}

\subsubsection{Normal mode splitting in a ring cavity pumped in one or both directions} \label{sec:BunchingNormalRing} 

Fig.\,\RefFig{Ring_Bunching} shows similar spectra as Fig.\,\RefFig{Linear_Bunching}, but for a ring cavity laser-pumped in one or both directions. The spectra are calculated from Eq.\,\RefEqg{Bunching08} using the ODM. 
 	\Fig{8.7}{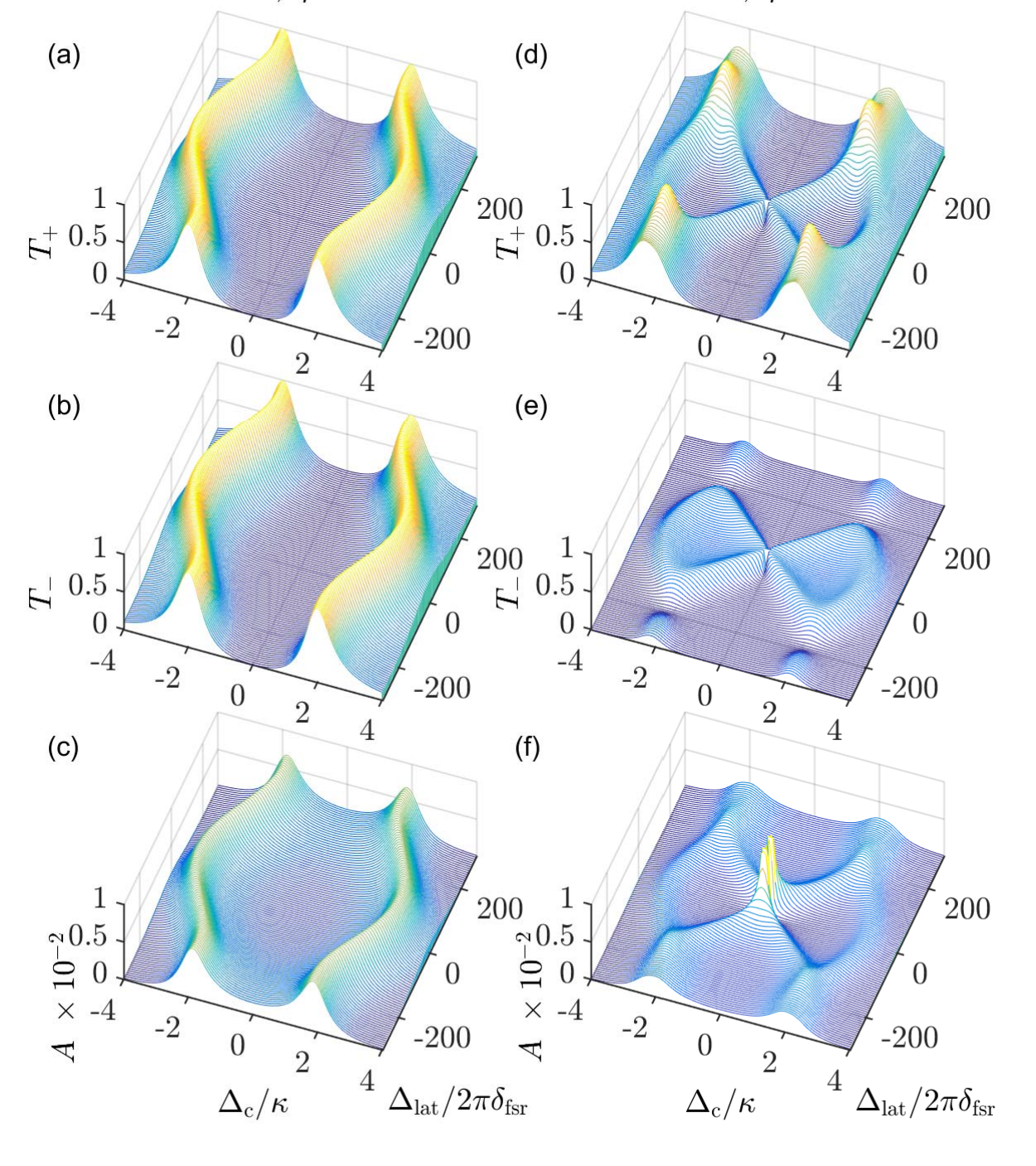}
		{Same as Fig.\,\RefFig{Linear_Bunching} but for (a-c) a ring cavity symmetrically pumped from both sides, $\eta_-=\eta_+$,
			with the atoms sitting at nodes when $\Delta_\text{lat}=0$, and (d-f) for a ring cavity pumped from one side only, $\eta_-=0$.
			The absorption is calculated as $A=1-T_+-T_--R_+-R_-$. Again, we stress that the TMM delivers identical solutions via 
			Eqs.\,\RefEqg{Bandgap64} and Eqs.\,\RefEqg{Bandgap66}.}

The symmetric pumping case closely reproduces the situation of a linear cavity. The panels Figs.\,\RefFig{Ring_Bunching}(a-c) are calculated for atoms that are perfectly bunched and all located at antinodes of the optical lattice. The spectra directly compare to Figs.\,\RefFig{Linear_Bunching}(a-c). For perfectly bunched atoms all located at nodes (not shown), we obtain spectra similar to those shown in Figs.\,\RefFig{Linear_Bunching}(d-f).

\bigskip

In contrast, the normal mode spectra in the unidirectionally pumped ring cavity deserve some extra discussion. We again start from Eq.\,\RefEqg{Bunching08}. For simplicity we neglect spontaneous emission, $\gamma_0=0$, in the remaining part of this section, which is justified in the bad cavity limit. In the case of uniform atomic distribution, $b=0$, the expression simplifies to,\Eqn
    {Bunching17}{\alpha_+ = \frac{\imath\eta_+}{\Delta_\text{c}-NU_0+\imath\kappa} ~~~\text{and}~~~ \alpha_- = 0~.}
This expression predicts standard normal mode splitting with two peaks and has been used, e.g.~in \cite{Culver16}. In Figs.\,\RefFig{Ring_Bunching}(d-f) this corresponds to $|\Delta_\text{lat}|\rightarrow\infty$, which is the regime where the lattice is totally incommensurate with the probe mode, so that the atoms can be considered as complete debunched. We observe peaks that, in resonance ($\Delta_\text{ca}\equiv\Delta_\text{c}-\Delta_\text{a}=0$), are located at $|\Delta_\text{c}|=g\sqrt{N}$.

\bigskip

Let us now concentrate on perfect bunching, $b_\pm=e^{\pm 2\imath kz_j}=e^{\pm 2\imath kz_0}$, and again look at the expression~\RefEqg{Bunching08} for the case one-sided pumping, $\eta_-=0$. In this case, the transmission profiles become more complicated, exhibiting up to four peaks. To understand the physical origin of the additional central peak near $\Delta_\text{c}=0$ in Figs.\,\RefFig{Ring_Bunching}(d-f), we combine the two counter-propagating modes \RefEqg{Bunching08} to a symmetric and an antisymmetric one, respectively,\Aln
    {Bunching18}{\alpha_++\alpha_-e^{-2\imath kz_0} & = \frac{\imath\eta_+\Delta_\kappa}{(\Delta_\kappa-NU_0)^2-N^2U_0^2}\\
		\alpha_+-\alpha_-e^{-2\imath kz_0} & = \frac{\imath\eta_+(\Delta_\kappa-2NU_0)}{(\Delta_\kappa-NU_0)^2-N^2U_0^2}~.\nonumber}
If the probe laser is kept in resonance with the cavity, $\Delta_\text{ca}=0$, the symmetric mode fully couples to the atoms thus generating maximum normal mode splitting, while the antisymmetric mode does not couple to the atoms, and the corresponding normal mode is not split. We observe peaks that in resonance are located at $|\Delta_\text{c}|=g\sqrt{N}$. For imperfect bunching ($\Delta_\text{lat}\ne 0$) the symmetric and antisymmetric mode get mixed, and both exhibit a certain amount of normal mode splitting, which explains the appearance of four normal modes in Fig.\,\RefFig{Ring_Bunching}(d-f) in the region of intermediate debunching $|\Delta_\text{lat}|/(2\pi\delta_\text{fsr})=0...150$. At large debunching the disorder is such, that the distinction between symmetric and anti-symmetric modes becomes meaningless. In this sense, the spectrum in Fig.\,\RefFig{Ring_Bunching}(d) can be understood as linear combination of spectra such as those shown in Figs.\,\RefFig{Linear_Bunching}(a) and (d).

\subsubsection{Avoided crossing curves} \label{sec:BunchingNormalAvoided} 

Plotting normal mode spectra for a unidirectionally pumped ring cavity as a function of $\Delta_\text{ca}$ and $\Delta_\text{a}$, we obtain the Figs.\,\RefFig{Ring_Avoided}. The panels (d-f) correspond to a totally disordered atomic cloud and show normal mode spectra similar to the ones observed in \cite{Rivero23} for the same set of parameters. However, as we assume low saturation in our present model, we are missing the bistable features observed in that work. A comparison between Figs.\,\RefFig{Ring_Avoided}(b) and (e) reveals that, if the atoms are not bunched, only a very small amount of light is back-scattered, while for high bunching Bragg reflection from the lattice generates a large backscattered amplitude. Indeed, in that work the atoms were not arranged in a lattice, but distributed along a running wave optical dipole trap generated by a cavity mode, so that the resonant ridges seen in Fig.\,\RefFig{Ring_Avoided}(e) were too weak to discern. 
 	\Fig{8.7}{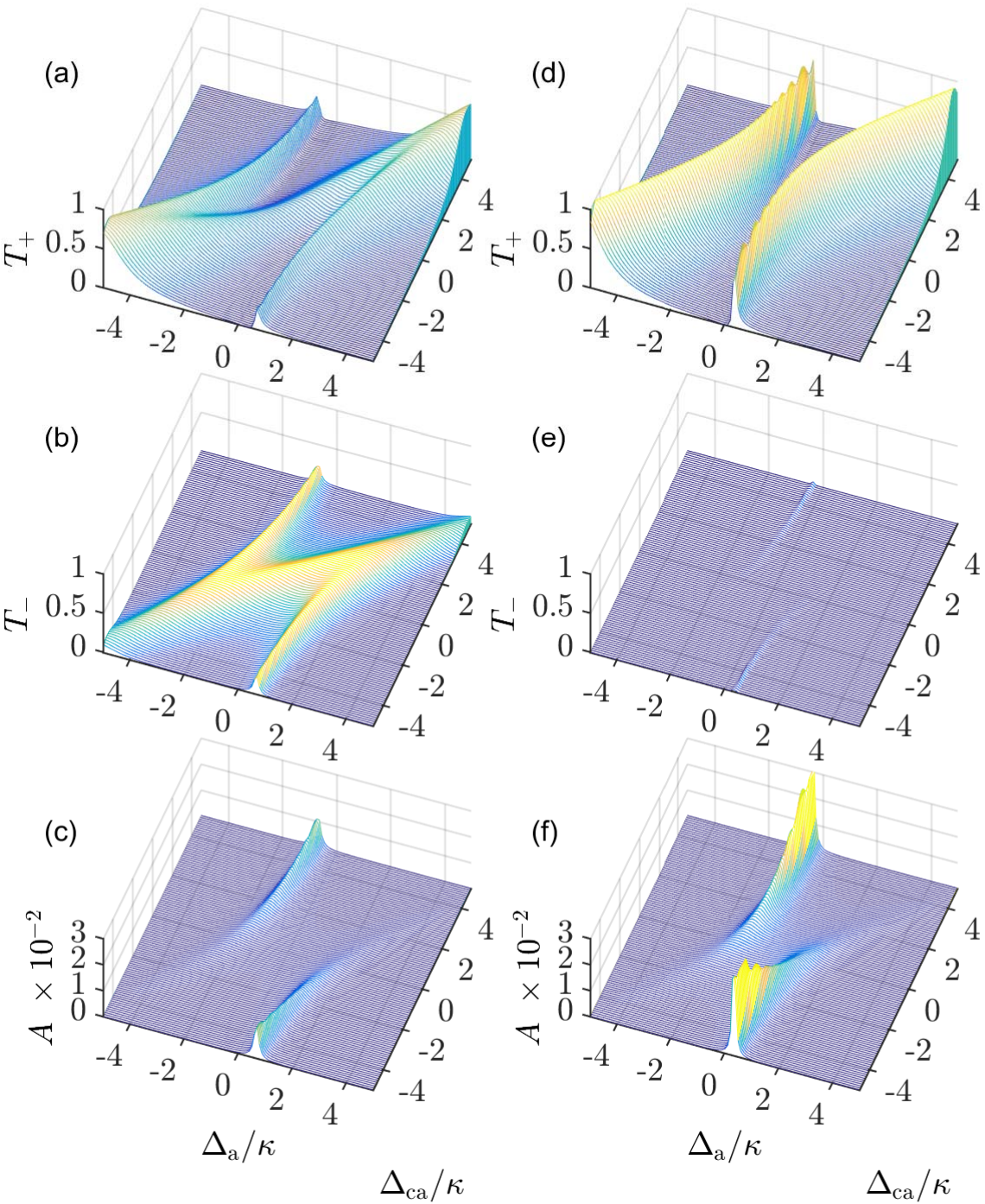}{Normal mode spectra for a ring cavity.
		Same as Fig.\,\RefFig{Ring_Bunching}, but now plotted as a function of laser-atom detuning 
		$\Delta_\text{a}\equiv\omega-\omega_\text{a}$ and atom-cavity detuning $\Delta_\text{ca}\equiv\omega_\text{a}-\omega_\text{c}$.
		(a-c) correspond to perfect bunching, $\Delta_\text{lat}=0$, and (d-f) to complete disorder. Same parameters as for
		 Fig.\,\RefFig{Ring_Bunching} except for $\Delta_\text{lat}=200\times 2\pi\delta_\text{fsr}$ and $N=2\times 10^5$. 
		Note that the same curves are obtained from the TMM.}

\subsection{Detection of Bloch oscillations in the tight-binding regime with coupling to a cavity} \label{sec:BunchingBlochoscis}

The dependence of the normal mode splitting on the number of atoms coupled to the cavity mode function can be exploited as a monitor for the time evolution of the atomic distribution along the cavity's optical axis. While this seems obvious for a linear cavity, whose mode function is spatially modulated, it is less so for a ring cavity. In this section we will show how to probe the atomic distribution in a uni-directionally pumped ring cavity via Bragg reflection.

When subjected to an external force, ultracold atoms confined in a lattice undergo Bloch oscillations \cite{Peik97}. A particularly interesting regime is the tight-binding limit \cite{Hartmann04,ZhangH23}, where the atoms coherently tunnel to neighboring lattice sites periodically spreading and refocussing their wavepackets. Cavities have been proposed to monitor non-destructively the oscillatory dynamics \cite{Peden09,Samoylova14,Samoylova15,ZhangH23}. By selecting a lattice period that is non-commensurate with the wavelength of the light probing the cavity, as illustrated in Fig.\,\RefFig{WannierStark}, the degree of atomic bunching can be made to depend on the extent of wavepacket spreading.
 	\Fig{7.7}{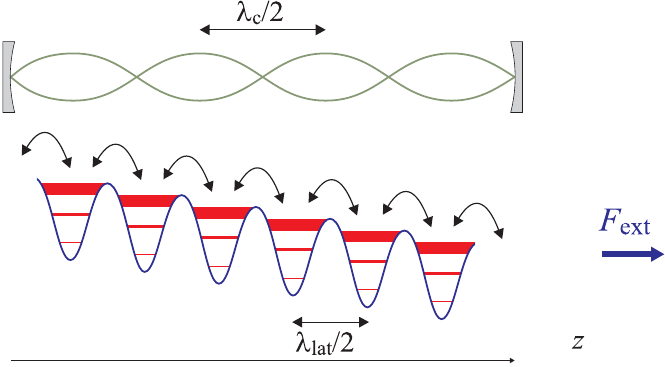}{Illustration of the Wannier-Bloch ladder in a lattice which is non-commensurate with the probe wavelength.
		Under the action of an external force $F_\text{ext}$, atoms in every lattice site will temporarily spread out to adjacent sites.}

The Wannier-Stark states, which are the eigenstates of the periodic potential containing the periodic potential plus the external force field, are orthonormal \cite{Hartmann04}. However, this breaks down in the presence of a cavity. After spreading, only the overlap between the Wannier-Stark states and the cavity mode function couples to the cavity. Let us consider an initially bunched distribution of uncorrelated atoms along the optical axis of the cavity, so that the expansion of the Wannier-Stark states in the Wannier state basis initially reads $|\psi_n(0)\>=\sum_jc_j(0)|n,j\>$, where $|n,j\>$ is the Wannier state labeled by the $n^\text{th}$ Bloch band and the $j^\text{th}$ lattice site and $c_j(0)\propto\cos^2(jk\lambda_\text{lat}/2)$, where $\lambda_\text{lat}/2$ is the periodicity of the lattice in which the atoms perform their Bloch oscillations. The Bloch oscillations correspond to a coherent evolution described by,\Eqn
	{Bunching20}{|\psi_n(t)\> = U_{jj'}(t)|\psi_n(0)\>~,}
where \cite{Hartmann04},\Eqn
	{Bunching21}{U_{jj'}(t) = e^{\imath(j-j')(\pi-\omega_\text{blo}t)/2-\imath j'\omega_\text{blo}t}J_{j-j'}(2\nu\sin\tfrac{\omega_\text{blo}t}{2})}
is the unitary evolution operator generating the Wannier-Bloch oscillations and $J_{j-j'}$ denotes Bessel functions. $\omega_\text{blo}=\frac{F_\text{ext}\lambda_\text{lat}}{2\hbar}$ is the Bloch oscillation frequency.

Fig.\,\RefFig{Ring_Bloch}(a) shows the periodic spreading and refocusing of one atom (or several atoms) initially located at a single lattice site labeled $j=0$. The maximum spreading $\nu$ (in units of numbers of lattice sites $j$) is given by the ratio between acceleration force $F_\text{ext}$ and tunneling rate, which can be calculated from the energy spectrum of the optical lattice \cite{Hartmann04}. In a homogeneously populated lattice, Wannier-Bloch oscillations do not change the numbers $|c_j(t)|^2$ of atoms in each lattice site, because loss and gain of atoms are balanced for every individual lattice. On the other hand, if the lattice sites initially contain different atom numbers, the Wannier-Bloch oscillations can lead to a time-dependent normalized population $|c_j(t)|^2$. As an example, Fig.\,\RefFig{Ring_Bloch}(b) shows a situation in which every forth lattice site between $j=-40$ and $j=40$ contains a number of $4N/N_\text{s}$ atoms, while $|c_j(0)|^2=0$ for the other sites.

While in free space, the distribution of atoms over lattice sites is normalized to the total atom number,\Eqn
	{Bunching22}{N = \sum_{j=(1-N_\text{s})/2}^{(N_\text{s}-1)/2}|c_j(t)|^2 ~~~\text{with}~~~ c_j(t) \equiv \<n,z_j|\psi_n(t)\>~.}
in a cavity the effective atom number contributing to normal mode splitting depends on the overlap with the cavity's mode function,\Eqn
	{Bunching23}{N_\text{eff}(t) = Nb_\mu(t) ~~~,~~~ \mu = 0,\pm~.}
Hence, within the ODM the impact of Wannier-Bloch oscillation is resumed in a time-dependent bunching parameter, which generalized from \RefEqg{Bunching11} becomes for a linear cavity,\Eqn
	{Bunching24}{b_0 = \frac{1}{N}\sum_{j=(1-N_\text{s})/2}^{(N_\text{s}-1)/2}|c_j(t)|^2\cos^2(jk\lambda_\text{lat}/2+kz_0)}
respectively, generalized from \RefEqg{Bunching12} for a ring cavity,\Eqn
	{Bunching25}{b_\pm = \frac{1}{N}\sum_{j=(1-N_\text{s})/2}^{(N_\text{s}-1)/2}|c_j(t)|^2e^{\pm 2\imath(jk\lambda_\text{lat}/2+kz_0)}~.}
Figs.\,\RefFig{Ring_Bloch}(c,d) show the variation in time of the bunching parameter as the atoms undergo the Wannier-Bloch oscillations shown Figs.\,(a,b).

Following the proposal of Ref.\,\cite{ZhangH23}, this time-varying bunching can be detected by monitoring the normal mode splitting of the transmission spectrum of a linear cavity.
 	\Fig{8.7}{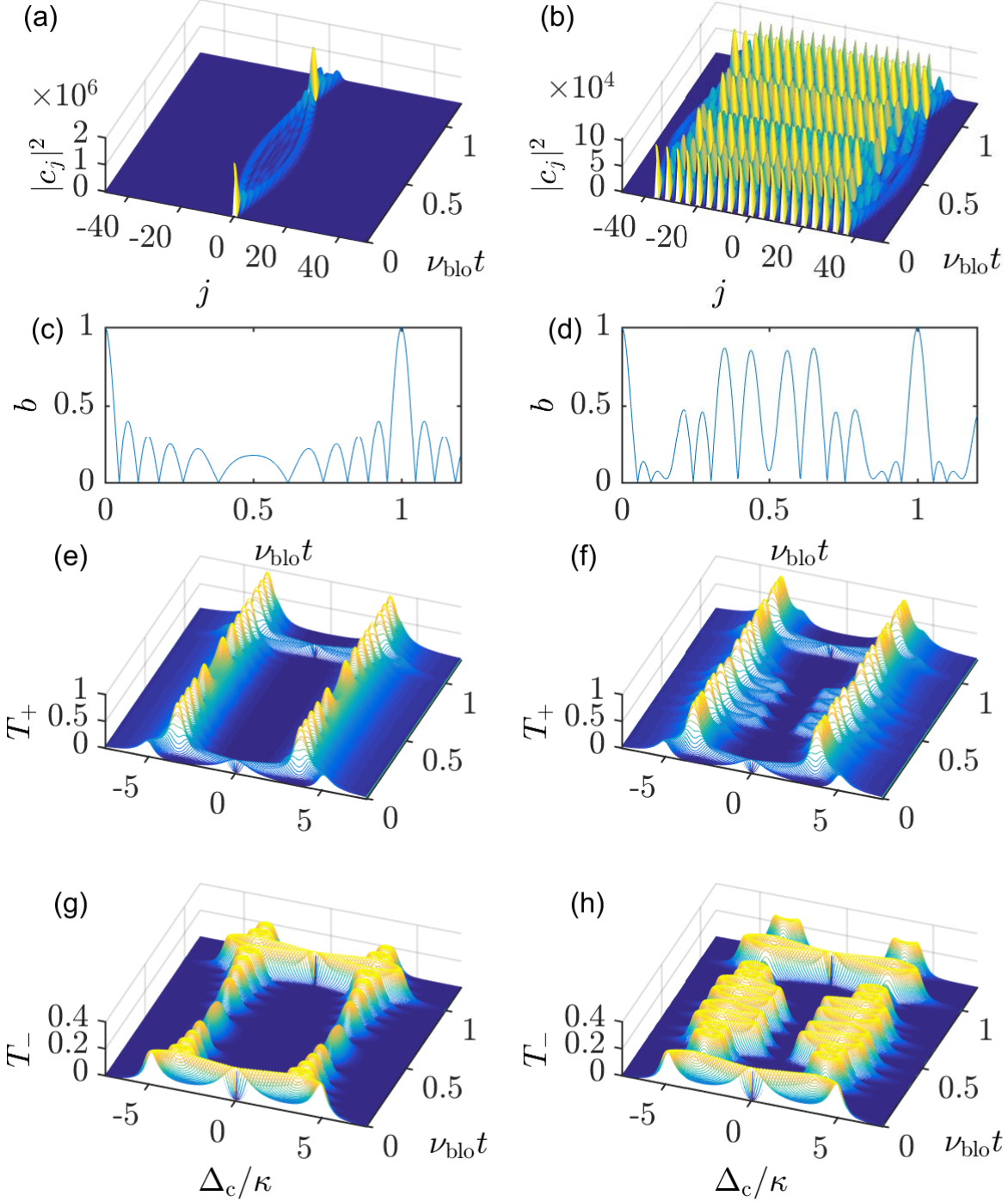}{Wannier-Bloch oscillations of atoms in a deep lattice. (a,c,e,g)~The cloud is initially located 
		in a single lattice in the center of the optical lattice $j=0$. (b,d,f,h)~The cloud is equally distributed over 
		$N_\text{s}=80$ lattice sites separated by $\Delta j=2$ and centered at $j=0$. Panels (a,b) show the spreading and 
		refocussing of the atom distribution over lattice sites upon Wannier-Bloch oscillations with $\nu=8$.
		(c,d)~show the corresponding bunching parameters, and (e,f) and (g,h)~show the normal mode transmission spectra $T_\pm$ 
		for a ring cavity varying over time. $N=2\cdot 10^6$, $\Delta_\text{lat}=0=\Delta_\text{ca}$. Other parameters as specified in 
		Sec.\,\ref{sec:BunchingNormalParameters}. The same curves are obtained from the TMM.}

In a ring cavity the mode functions are translation-invariant so that, when the cavity is pumped from one side, one might expect that the atomic positions should have no impact on normal mode splitting. This is indeed true for the transmission signal $T_+$, as seen in Figs.\,\RefFig{Ring_Bloch}(e,f). However, atomic bunching engenders coupling between the counter-propagating cavity modes via Bragg reflection. The pumped mode $\alpha_+$ interacts with the reverse mode $\alpha_-$ at a coupling strength given by $Nb_\mu U_0$. Consequently, if the probe light frequency is sufficiently close to a normal mode, at any instants of time in which the Wannier-Bloch oscillations generate bunching, light is back-scattered from the pumped mode $\alpha_+$ into the mode $\alpha_-$. The interaction of this back-scattered light with the atoms generates normal mode splitting which can be observed in the signal $T_-$, as shown in Figs.\,\RefFig{Ring_Bloch}(g,h). When the probe light is detuned from the normal modes, the coupling only leads to a phase shift of the mode $\alpha_+$, which can be detected by homodyne techniques.

\bigskip

In practice, the atomic cloud could be pre-bunched in a deep lattice with periodicity $\lambda_\text{lat}=\frac{m}{n}\lambda$, where $m,n\in\mathbb{Z}$ are small integer numbers. The lattice depth is then adjusted such that the atoms perform Bloch oscillations in the tight-binging regime. The reflection of a  probe laser injected into one ring cavity mode then monitors the normal mode splitting.

\section{Transfer matrix model} \label{sec:Transfer}

All calculations and simulations presented so far were derived from the open Dicke model (ODM) based on the assumption that the atomic cloud is optically dilute, i.e.~has low optical density ($OD<1$), and that the intracavity light intensity does not saturate the atomic transition, i.e.~$|\alpha|<\Gamma/2g$. For the parameters used (small enough atom numbers $N$ and lattice sites $N_\text{s}$ and highly reflecting mirrors, $R_\text{mir}=99.8\%$ being the mean reflectivity of all mirrors) both the ODM and the TMM yield identical results. That is, all graphs generated so far are perfectly reproduced by the TMM. However, outside this parameter regime, two models begin to diverge.

Indeed, for high optical densities $OD\gtrsim 1$, such as those achieved in the experiment of Ref.\,\cite{Rivero23}, the ODM reaches its limitations. In this regime, other models are needed that can account in a simple way for propagation effects and effects due to a locally varying refractive index. The Transfer Matrix Model (TMM) is one such model. Before we present and discuss the differences, let us briefly recapitulate the transfer matrix model and extend it to the cases of a one-dimensional optical lattice aligned with the optical axis of a linear and of a ring cavity.

\subsection{Link between atom-cavity coupling constant and single-atom reflection} \label{sec:TransferPolarizability}

The TMM describes the variation of the electric field amplitudes in both counter-propagating directions along the optical axis and through optical components or layers of atomic scatterers \cite{Hemmerich99}. Based on the complex atomic polarizability,\Eqn
	{Bandgap03}{\frac{\alpha_\text{pol}}{\varepsilon_0} \simeq \frac{6\pi}{k^3}\frac{-1}{\imath+2\Delta_\text{a}/\Gamma}~,}
the \textit{single atom reflection coefficient} is defined as,\Eqn
	{Bandgap04}{\beta_\Delta = \frac{k}{\pi w^2}\frac{\alpha_\text{pol}}{\varepsilon_0}
		= \frac{6}{k^2w^2}\frac{-1}{\imath+2\Delta_\text{a}/\Gamma}~.}
The resonant reflection coefficient describes how well the resonant optical cross-section of the atom, $\sigma_0=3\lambda^2/2\pi$, matches the cross section of the optical mode, $\pi w^2$,\Eqn
	{Bandgap05}{-\imath\beta_0 = \frac{\sigma_0}{\pi w^2} = \frac{6}{k^2w^2} = \frac{4g^2}{\delta_\text{fsr}\Gamma} = \frac{\pi\Upsilon}{F}~.}
The reflection coefficient multiplied with the free spectral range of the cavity,\Aln
	{Bandgap06}{\delta_\text{fsr}\beta_\Delta & = \frac{6\delta_\text{fsr}}{k^2w^2}\frac{-1}{\imath+2\Delta_\text{a}/\Gamma}\\
		& = g^2\frac{\Delta_\text{a}-\imath\Gamma/2}{\Delta_\text{a}^2+\Gamma^2/4} = U_0-\imath\gamma_0 = U_\gamma~,\nonumber}
is just the \emph{single-photon light-shift} combined with the single-photon Rayleigh scattering rate introduced in Eq.\,\RefEqg{Bunching04}. 

In the presence of many atoms the atom-field coupling is collectively enhanced, and we may define a collective atom-field coupling constant as $g_N\equiv g\sqrt{N}$. The phase shift caused by $N$ atoms is then $N\beta_\Delta$. As we are interested in the atomic density per lattice period, we replace $N$ by the number of atoms $N_1$ in each one of the $N_\text{s}$ lattice sites. Note that, as our model is strictly one-dimensional, the radial distribution of the atoms in the optical lattice (whose mode function is assumed to be the same as the one of the probe laser) does not matter. Hence, the reflectivity of a single layer containing $N_1=N/N_\text{s}$ atoms is simply given by,\Eqn
    {Bandgap07}{\beta_1 \equiv N_1\beta_\Delta~.}
Typical experiments \cite{Rivero23} involve $N\approx 200000$ atoms distributed over $N_\text{s}\approx 300$ antinodes of the standing light wave potential. With the beam waist $w$ specified in Sec.\,\ref{sec:BunchingNormalParameters} and the lattice wavelength $\lambda_\text{lat}=689\U{nm}$, and supposing the atoms to be radially homogeneously distributed over the beam waist, this corresponds to an average density of $n_\text{at}\approx 10^{11}\U{cm^{-3}}$. The resonant optical density of the cloud in the cavity is given by the ratio between the collective cooperativity $\Upsilon_N$ and the finesse $F$,\Eqn
	{Bandgap08}{OD = \sigma_0n_\text{at}N_\text{s}\frac{\lambda_\text{lat}}{2} = \frac{6N}{k^2w^2} = \frac{\pi\Upsilon_N}{F} \approx 3~.}

\bigskip

\emph{Strong coupling} and \emph{high optical density} are obviously different concepts. A thin slab of matter may have a high refraction index, $|n_\text{rfr}-1|\gg 0$, leading to strong coupling, but it can still be optically dilute with $OD<1$. Eq.\,\RefEqg{Bandgap08} tells us that in a cavity characterized by $F\gg 1$ a high optical density implies strong coupling, but this is not always the case \cite{Mahmoodian18,PrasadA20}.

\subsection{Transfer matrix model for normal-mode splitting} \label{sec:TransferTransfer}
 
The goal of the following derivations is to show that, applied to the coupled system of a cavity interacting with a cloud of atoms, the TMM not only reproduces the well-known normal splitting, but also conveniently allows to calculate features arising from atomic order or disorder at high optical densities. The TMM extends beyond the capabilities of the ODM by including the aspects of light-mediated interatomic interactions related to the distance between the atoms \cite{Samoylova14,Samoylova14b}. In particular, one-dimensional photonic bandgaps are conveniently described within the transfer matrix formalism \cite{Deutsch95b,Slama06,Schilke11}.

To prepare the ground for the TMM, let us derive the relevant transfer matrices for our coupled atom-cavity system. We proceed by steps recapitulating the formalism for (i)~the reflection and transmission of an empty linear cavity, (ii)~a linear cavity containing a 1D coaxial optical lattice, (iii)~the intensity distribution inside the linear cavity, and (iv)~a ring cavity with two coupled counter-propagating modes.

\subsubsection{Airy formula from the TMM for an empty linear cavity} \label{sec:TransferTransferAiry}

A \emph{transfer matrix} $\mathcal{T}$ transforms a pair of field amplitudes belonging to counter-propagating modes known at one point (1) of the optical axis into a pair at point (2) according to,\begin{small}\Eqn
    {Bandgap21}{\Mtz{\alpha_+^{(2)} \\ \alpha_-^{(2)}} = \mathcal{T}_{1\rightarrow 2}\Mtz{\alpha_+^{(1)} \\ \alpha_-^{(1)}} ~~~\text{with}~~~ 
		\mathcal{T}_{1\rightarrow 2} = \Mtz{T^{11} & T^{12} \\ T^{21} & T_{22}}~.}\end{small}
The corresponding \emph{scattering matrix} $\mathcal{S}$ is defined by,\begin{small}\Eqn
    {Bandgap22}{\Mtz{\alpha_+^{(2)} \\ \alpha_-^{(1)}} = \mathcal{S}_{1\leftrightarrow 2}\Mtz{\alpha_+^{(1)} \\ \alpha_-^{(2)}} ~~~\text{with}~~~ 
		\mathcal{S}_{1\leftrightarrow 2} = \Mtz{S^{11} & S^{12} \\ S^{21} & S_{22}}}\end{small}
and related to the transfer matrix via partial inversion,\begin{small}\Eqn
    {Bandgap23}{\mathcal{S}_{1\leftrightarrow 2} = \tfrac{1}{T^{22}}\Mtz{T^{11}T^{22}-T^{12}T^{21} & T^{12} \\ -T^{21} & 1}~,}\end{small}
where $T^{ij}$ are the matrix elements of $\mathcal{T}_{1\rightarrow 2}$, respectively,\begin{small}\Eqn
    {Bandgap24}{\mathcal{T}_{1\rightarrow 2} = \tfrac{1}{S^{22}}\Mtz{S^{11}S^{22}-S^{12}S^{21} & S^{12} \\ -S^{21} & 1}~,}\end{small}
where $S^{ij}$ are the matrix elements of $\mathcal{S}_{1\leftrightarrow 2}$.
	
\bigskip

Let us now consider an empty linear cavity of length $L$. With the free spectral range $\delta_\text{fsr}\equiv c/2L$, we can write the wavevector of the incident probe light as,\begin{small}\Eqn
    {Bandgap22}{k = \frac{\omega}{c} = \frac{\omega_\text{a}+\Delta_\text{a}}{c}
        = \frac{\omega_\text{c}+\Delta_\text{c}}{c} = \frac{N_\text{mod}2\pi\delta_\text{fsr}+\Delta_\text{c}}{c}}\end{small}
with $N_\text{mod}\in\mathbb{N}$. The transfer matrix for free space propagation over a distance $z$ located somewhere between the cavity mirrors then reads,\Eqn
    {Bandgap26}{\boxed{\mathcal{P}_z(\Delta_\text{c}) = \Mtz{e^{\imath kz} & 0 \\ 0 & e^{-\imath kz}}}~.}
Hence, $\mathcal{P}_L(\Delta_\text{c})$ is the propagation matrix between the cavity mirrors with $e^{\imath kL}=e^{\imath\Delta_\text{c}/2\delta_\text{fsr}}$. 
 	\Fig{8.7}{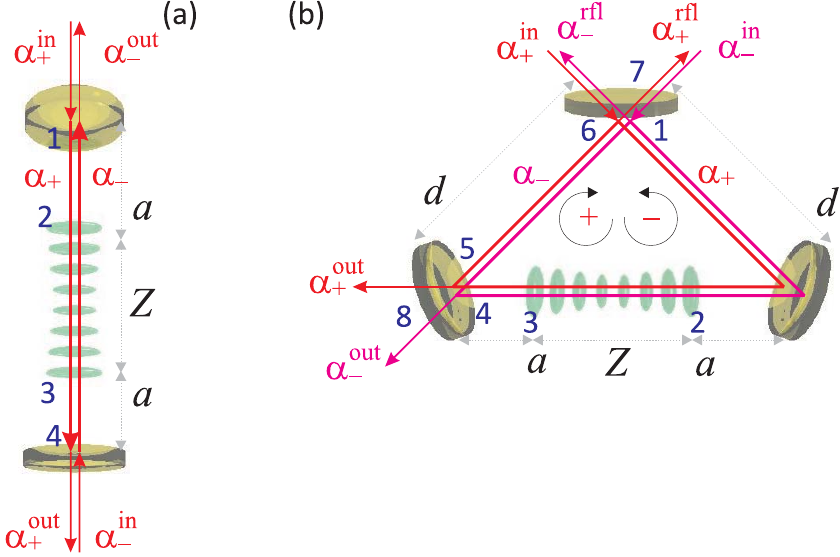}{Same scheme as in Fig.\,\RefFig{Scheme1} but adapted to illustrate the TMM applied to (a)~a linear cavity and (b)~a ring cavity. 
		Blue numbers refer to specific positions $z$ on the optical axis, where the electric field $\alpha_\pm^{(z)}$ is evaluated (see text). 
		Positive (negative) subscripts correspond to clockwise (counter-clockwise) propagation.}

\bigskip

The \emph{scattering matrix} for a beam splitter with transmissivity $t_\text{bs}$ and reflectivity $r_\text{bs}$ is,\Eqn
    {Bandgap27}{\mathcal{S}_\text{bs} = \Mtz{t_\text{bs} & -r_\text{bs} \\ r_\text{bs} & t_\text{bs}}}
when absorption losses can be neglected, in which case, $\det\mathcal{S}_\text{bs}=1$. Applied to the first input coupling mirror of the linear cavity sketched in Fig.\,\RefFig{Scheme2}(a),\Eqn
    {Bandgap28}{\Mtz{\alpha_+^{(1)} \\ \alpha_-^\text{out}} = \mathcal{S}_\text{bs}\Mtz{\alpha_+^\text{in} \\ \alpha_-^{(1)}}}
Using the prescription \RefEqg{Bandgap24} we obtain from \RefEqg{Bandgap27} the \emph{transfer matrix},\Eqn
    {Bandgap29}{\boxed{\mathcal{T}_\text{bs} = \tfrac{1}{t_\text{bs}}\Mtz{1 & -r_\text{bs} \\ -r_\text{bs} & 1}}~,}
For the mirror at the position (1) ($z=0$) and the second mirror at (4) ($z=L$), we thus get [see labeling introduced in Fig.\,\RefFig{Scheme2}(a)],\begin{small}\Eqn
    {Bandgap30}{\Mtz{\alpha_+^{(1)} \\ \alpha_-^{(1)}} = \mathcal{T}_\text{bs1}\Mtz{\alpha_+^\text{in} \\ \alpha_-^\text{out}} ~~\text{and}~~
		\Mtz{\alpha_+^{(4)} \\ \alpha_-^{(4)}} = \mathcal{T}_\text{bs2}\Mtz{\alpha_+^\text{out} \\ \alpha_-^\text{in}}}\end{small}
where $\alpha_+^\text{in}=\eta/\kappa$ and $\alpha_-^\text{in}=0$. Intensity losses (as they may occur e.g.~upon passage through additional optical components or reflection at additional cavity mirrors with reflectivity $r_\text{ls}$), are described by,\Eqn
    {Bandgap31}{\boxed{\mathcal{T}_\text{ls} = \Mtz{\pm r_\text{ls} & 0 \\ 0 & \pm r_\text{ls}^{-1}}}~,}
where the negative signs account for $180^\circ$ phase shifts upon reflections.

The total transfer matrix of the empty cavity is obtained by concatenation,\Eqn
    {Bandgap32}{\mathcal{T}_\text{tot} = \mathcal{T}_\text{bs2}^{-1}~\mathcal{P}_\text{L}(\Delta_\text{c})~\mathcal{T}_\text{bs1}~.}
Finally, in order to express the output as a function of the input signals, we reconvert to the scattering matrix of the cavity as a whole,\Eqn
    {Bandgap33}{\boxed{\Mtz{\alpha_+^\text{out} \\ \alpha_-^\text{out}} = \mathcal{S}_\text{tot}\Mtz{\alpha_+^\text{in} \\ \alpha_-^\text{in}}}}
exploiting the prescription \RefEqg{Bandgap23}. The transmission spectra are now obtained by setting $\alpha_-^\text{in}=0$ in \RefEqg{Bandgap33} and calculating,\Eqn
    {Bandgap34}{T = \left|\frac{\alpha_+^\text{out}(\Delta_\text{c})}{\alpha_+^\text{in}}\right|^2~,}
The transfer matrix \RefEqg{Bandgap32}, and hence the scattering matrix in \RefEqg{Bandgap33}, only depend on experimental parameters and allow us to calculate the response of the atom-cavity system to any incident field. An analytical calculation of $T$ with the given transfer matrices reproduces the well-known Airy formula for the empty cavity.

\subsubsection{Transmission from the TMM for a cavity containing an optical lattice} \label{sec:TransferTransferPhotonic}

Now, we consider a cloud of atoms with resonance frequency $\omega_\text{a}$ trapped inside a standing light wave potential tuned to a frequency $\omega_\text{lat}$. The atoms interact with a linear optical cavity, whose nearest mode is at the frequency $\omega_\text{c}$. The cavity mode is collinearly pumped by a probe laser frequency $\omega$. The transfer matrix for the atomic cloud is expressed in terms of the single atomic layer reflectivity derived in \RefEqg{Bandgap07},\Eqn
    {Bandgap35}{\boxed{\mathcal{A} = \Mtz{1+\imath\beta_1 & \imath\beta_1 \\ -\imath\beta_1 & 1-\imath\beta_1}}~.}
The total transfer matrix for the atomic cloud is then,\Eqn
    {Bandgap36}{\mathcal{A}_\text{tot}(\Delta_\text{c}) = [\mathcal{A}\,\mathcal{P}_{\lambda_\text{lat}/2}(\Delta_\text{c})]^{N_\text{s}}~.}
Now, the complete total transfer matrix for a linear cavity containing an optical lattice can be obtained simply by extending Eq.\,\RefEqg{Bandgap32},\Eqn
    {Bandgap37}{\mathcal{T}_\text{tot} = \mathcal{T}_\text{bs2}^{-1}\,\mathcal{P}_{a}(\Delta_\text{c})\,
		\mathcal{A}_\text{tot}(\Delta_\text{c})\,\mathcal{P}_{a}(\Delta_\text{c})\,\mathcal{T}_\text{bs1}~,}
where $a$ is the distance between the optical lattice and the input coupling mirror [see Fig.\,\RefFig{Scheme2}(a)].

\bigskip

As already mentioned, all graphs obtained from the ODM and exhibited in Sec.\,\ref{sec:Bunching} are perfectly reproduced by the TMM, provided the optical density of the atomic cloud is low. Deviations are observed in the presence of strong \emph{absorption} or \emph{reflection}, where 'strong' means that the light beam suffers noticeable attenuation along its path through the cavity. Figs.\,\RefFig{Linear_Ring}(a,b) compare transmission spectra obtained from the ODM and the TMM for the same parameters. To work out the differences of both models, the coupling constant $g$ has been increased and the mean cavity mirror reflectivity decreased. Apparently, the TMM predicts additional resonances at smaller detunings $\Delta_\text{c}/\Gamma$. Increasing $g$ enhances the optical density, but the role the mirror reflectivity will only be unraveled in Sec.\,\ref{sec:BandgapPhotonic}.
 	\Fig{8.7}{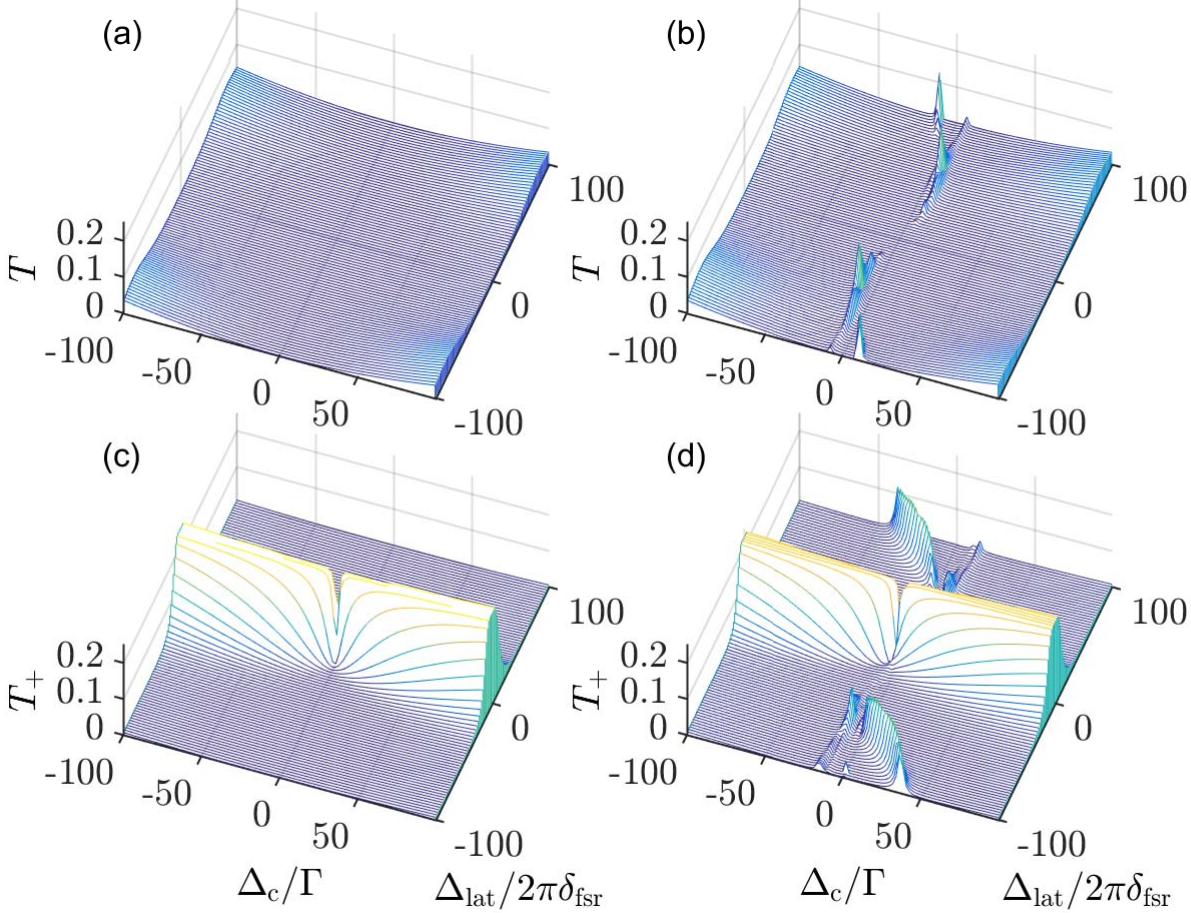}{Transmission spectra as a function of laser detuning and lattice constant calculated for 
		(a,b)~ a linear cavity and (c,d)~a ring cavity from (a,c)~the ODM and (b,d)~the TMM.
		The mean reflectivity of the cavity mirror is reduced to $90\%$ and the coupling constant increased to $g=10\Gamma$.
		The slight increase of transmission in the four corners of (a,b) comes from normal mode splitting.}

\subsubsection{Intensity inside the cavity} \label{sec:TransferTransferIntra}

To calculate the intensity distribution along the optical axis inside the cavity, we assume that the optical lattice is located between the points $(2)$ and $(3)$ of the optical axis, as indicated in Fig.\,\RefFig{Scheme2}(a). The transfer through the entire structure (linear cavity + lattice) up to a point $z$ is expressed as,\Eqn
    {Bandgap41}{\Mtz{\alpha_+^{(z)} \\ \alpha_-^{(z)}} 
        = \mathcal{T}_\text{(z)}\Mtz{\alpha_+^\text{in} \\ \alpha_-^\text{out}}
	= \mathcal{X}_{(z)}\Mtz{\alpha_+^\text{in} \\ \alpha_-^\text{in}}~,}
where $\mathcal{T}_\text{(z)}$ is the transfer matrix concatenation containing the input coupler and all elements located between the input coupler and the position $z$ of the optical axis, and\Eqn
    {Bandgap42}{\mathcal{X}_{(z)} \equiv \mathcal{T}_{(z)}\left[\Mtz{1 & 0 \\ 0 & 0}+\Mtz{0 & 0 \\ 0 & 1}\mathcal{S}_\text{tot}\right]~.}

Using the information that the cavity is only pumped from one side, $\alpha_-^\text{in}=0$, we get explicitly,\Eqn
    {Bandgap43}{\Mtz{\alpha_+^{(z)} \\ \alpha_-^{(z)}} = \mathcal{T}_\text{(z)}\Mtz{1 \\ -\mathcal{T}^{21}/\mathcal{T}^{22}}\alpha_+^\text{in}~.}
The sum of the counter-propagating field amplitudes at position $z$ normalized by the incident field amplitude is,\Eqn
    {Bandgap44}{\boxed{\frac{\alpha_+^{(z)}+\alpha_-^{(z)}}{\alpha_+^\text{in}} 
		= \mathcal{T}^{11}_\text{(z)}+\mathcal{T}^{21}_\text{(z)}
			-\frac{\mathcal{T}_\text{tot}^{21}}{\mathcal{T}_\text{tot}^{22}}(\mathcal{T}^{12}_\text{(z)}+\mathcal{T}^{22}_\text{(z)})}~.}

\subsection{Transfer matrices for an optical lattice inside a ring cavity} \label{sec:TransferRing}

The transfer matrix formalism can be applied to ring cavities in an analogous fashion as for linear cavities. A difference is, however, that one has to deal with two independent counter-propagating modes, which are \emph{only coupled in the presence of atoms} scattering light between the modes.

\subsubsection{Transfer matrices for ring cavities} \label{sec:TransferRingMatrices}

To calculate the intensity distribution inside and behind a laser-pumped ring cavity, we use the $\mathcal{S}$-matrices for the incoupling beam splitter at point (7),\begin{small}\Eqn
    {Bandgap51}{\Mtz{\alpha_+^{(1)}\\
        \alpha_+^\text{rfl}} = \mathcal{S}_\text{ic}\Mtz{\alpha_+^\text{in} \\ \alpha_+^{(6)}} ~~\text{and}~~ 
			\Mtz{\alpha_-^{(6)} \\ \alpha_-^\text{rfl}} = \mathcal{S}_\text{ic}\Mtz{\alpha_-^\text{in} \\ \alpha_-^{(1)}}~,}\end{small}
and for the outcoupling beam splitter at point (8), as indicated in Fig.\,\RefFig{Scheme2}(b),\begin{small}\Eqn
    {Bandgap52}{\Mtz{\alpha_+^{(5)} \\ \alpha_+^\text{out}} = \mathcal{S}_\text{hr}\Mtz{0 \\ \alpha_+^{(4)}} ~~\text{and}~~ 
        \Mtz{\alpha_-^{(4)} \\ \alpha_-^\text{out}} = \mathcal{S}_\text{hr}\Mtz{0 \\ \alpha_-^{(5)}}}\end{small}
with the beam splitting $\mathcal{S}_\text{bs}$-matrix given in \RefEqg{Bandgap27}. 

The $\mathcal{T}$-matrices for propapation along the ring cavity's optical axis and across the optical lattice are the same $2\times 2$-matrices introduced for linear cavities: Eq.\,\RefEqg{Bandgap21} is the $\mathcal{T}$-matrix describing the clockwise transfer of a laser beam inside a ring cavity from point (1) to point (2), and Eqs.\,\RefEqg{Bandgap26} and \RefEqg{Bandgap35} respectively describe free-space propagation and reflection at atomic layers. Finally, the matrix \RefEqg{Bandgap31} describes losses at the third mirror of the ring cavity. The transfer matrix describing a complete round-trip through the ring cavity from point (1) to (6),\Eqn
	{Bandgap53}{\mathcal{T}_{1\rightarrow 6} \equiv \Mtz{R_{11} & R_{12} \\ R_{21} & R_{22}}~,}
is derived by concatenation of transfer matrices in the same way as demonstrated for the linear cavity in Eq.~\RefEqg{Bandgap37}. Complying with the notation of Fig.\,\RefFig{Scheme2}(b) we write,\Eqn
	{Bandgap54}{\Mtz{\alpha_+^{(6)} \\ \alpha_-^{(6)}} = \mathcal{T}_{1\rightarrow 6}\Mtz{\alpha_+^{(1)} \\ \alpha_-^{(1)}}~.}
Combining this with the Eq.\,\RefEqg{Bandgap51}, we obtain a system of six independent equations. Eliminating the field amplitudes at point (6) from these equations, we are left with,\begin{small}\Eqn
    {Bandgap55}{\Mtz{(1+r_\text{ic}R_{11})\alpha_+^{(1)}+r_\text{ic}R_{12}\alpha_-^{(1)} \\
        -t_\text{ic}R_{11}\alpha_+^{(1)}-t_\text{ic}R_{12}\alpha_-^{(1)}+\alpha_+^\text{rfl}\\
        R_{21}\alpha_+^{(1)}+(r_\text{ic}+R_{22})\alpha_-^{(1)}\\
        -t_\text{ic}\alpha_-^{(1)}+\alpha_-^\text{rfl}} = \Mtz{t_\text{ic}\alpha_+^\text{in} \\ r_\text{ic}\alpha_+^\text{in} \\ 
			t_\text{ic}\alpha_-^\text{in} \\r_\text{ic}\alpha_-^\text{in}}~,}\end{small}
The first and third equation yield,\Eqn
    {Bandgap56}{\mathcal{Y}^{-1}\Mtz{\alpha_+^{(1)} \\ \alpha_-^{(1)}} = \Mtz{\alpha_+^\text{in} \\ \alpha_-^\text{in}}}
with\Eqn
	{Bandgap57}{\mathcal{Y}^{-1} = \tfrac{1}{t_\text{ic}}\Mtz{1+r_\text{ic}R_{11} & r_\text{ic}R_{12} \\ R_{21} & r_\text{ic}+R_{22}}~,}
or, resolved by the intracavity field amplitudes,\Eqn
    {Bandgap59}{\Mtz{\alpha_+^{(1)} \\ \alpha_-^{(1)}} = \mathcal{Y}\Mtz{\alpha_+^\text{in} \\ \alpha_-^\text{in}}~.}
The matrix $\mathcal{Y}$ describes, how the field amplitudes at point (1) just behind the input coupler depend on both incident light fields $\alpha_\pm^\text{in}$. However, it is neither a transfer nor a scattering matrix, as it depends on all components inside the ring cavity via the coefficients $R_{ij}$ of the complete round-trip matrix $\mathcal{T}_{1\rightarrow 6}$.

\subsubsection{Intensity in- and outside the cavity} \label{sec:TransferRingIntra}

Now, beginning at point (1), we can express the field amplitudes at any point simply by concatenating $\mathcal{T}$-matrices. For an arbitrary location $z$, we have,\Eqn
    {Bandgap62}{\boxed{\Mtz{\alpha_+^{(z)} \\ \alpha_-^{(z)}} 
		= \mathcal{T}_{1\rightarrow z}~\mathcal{Y}\Mtz{\alpha_+^\text{in} \\ \alpha_-^\text{in}}}~.}
For the fields outcoupled at points (4) and (5) we get,\Eqn
    {Bandgap63}{\Mtz{\alpha_+^\text{out} \\ \alpha_-^\text{out}} = t_\text{hr}\Mtz{\alpha_+^{(4)} \\ \alpha_-^{(5)}}~.}
Hence,\Eqn
    {Bandgap64}{\boxed{\Mtz{\alpha_+^\text{out} \\ \alpha_-^\text{out}} 
		= \Mtz{t_\text{hr} & 0 \\ 0 & -\tfrac{t_\text{hr}}{r_\text{hr}}}\mathcal{T}_{1\rightarrow 4}~\mathcal{Y}\Mtz{\alpha_+^\text{in} \\ \alpha_-^\text{in}}}~.}

To calculate the reflected amplitudes we consider the second and forth equation \RefEqg{Bandgap55},\begin{small}\Eqn
    {Bandgap65}{-t_\text{ic}\Mtz{R_{11} & R_{12} \\ 0 & 1}\Mtz{\alpha_+^{(1)} \\ \alpha_-^{(1)}}+\Mtz{\alpha_+^\text{rfl} \\ \alpha_-^\text{rfl}}
		= r_\text{ic}\Mtz{\alpha_+^\text{in} \\ \alpha_-^\text{in}}}\end{small}
Solving by the reflected amplitudes,\Eqn
    {Bandgap66}{\boxed{\Mtz{\alpha_+^\text{rfl} \\ \alpha_-^\text{rfl}} 
        \equiv \mathcal{X}\Mtz{\alpha_+^\text{in} \\ \alpha_-^\text{in}}}}
with\begin{small}\Aln
    {Bandgap67}{\mathcal{X} & = r_\text{ic}\mathbb{I}+t_\text{ic}\Mtz{R_{11} & R_{12} \\ 0 & 1}\mathcal{Y}\\
        & = \tfrac{1}{D}\Mtz{(r_\text{ic}+R_{11})(r_\text{ic}+R_{22})-R_{12}R_{21} ~~~ t_\text{ic}^2R_{12} \\
			-t_\text{ic}^2R_{21} ~~~ (1+r_\text{ic}R_{11})(1+r_\text{ic}R_{22})-r_\text{ic}^2R_{12}R_{21}}\nonumber}\end{small}
and the abbreviation\Eqn
    {Bandgap68}{D \equiv \det\Mtz{r_\text{ic}+R_{22} & -r_\text{ic}R_{12} \\ -R_{21} & 1+r_\text{ic}R_{11}}~.}

Finally, to evaluate the fields \RefEqg{Bandgap62}, \RefEqg{Bandgap64}, and \RefEqg{Bandgap66}, we still need to calculate the components of the round trip matrix,\Eqn
    {Bandgap69}{\mathcal{T}_{1\rightarrow 6} = \mathcal{P}(d)\,\mathcal{T}_\text{ls}\,\mathcal{T}_{1\rightarrow 4}}
with\Eqn
    {Bandgap70}{\mathcal{T}_{1\rightarrow 4} = \mathcal{P}(a)\,\left[\mathcal{A}\,\mathcal{P}(\tfrac{\lambda_\text{lat}}{2})\right]^{N_\text{s}}
		\,\mathcal{P}(a)\,\mathcal{T}_\text{ls}\,\mathcal{P}(d)~,}
where the distances $a$ and $d$ are defined in the scheme of Fig.\,\RefFig{Scheme2}(b).

\bigskip

The expressions \RefEqg{Bandgap64} and \RefEqg{Bandgap66} allow us to calculate the transmission and reflection signals from the expressions \RefEqg{Bunching10} based on the TMM. For the parameters used to calculate the spectra of Figs.\,\RefFig{Ring_Bunching} and \RefFig{Ring_Avoided} based on the ODM, we reproduce exactly the same spectra with the TMM. Just as for the case of linear cavities, the two models deviate from each other as soon as the optical density is large. 
Figs.\,\RefFig{Linear_Ring}(c,d) compare transmission spectra obtained from the ODM and the TMM for the identical parameters. Similarly to the case of a linear cavity, the TMM predicts additional resonances at smaller detunings $\Delta_\text{c}/\Gamma$, which will be discussed in more detail in Sec.\,\ref{sec:BandgapPhotonic}.

\section{Photonic bands in cavities} \label{sec:Bandgap}

\subsection{Light propagation through a cavity} \label{sec:BandgapPropagation}

In order to understand better the impact of absorption and phase shifts caused by the optical lattice, we use our transfer matrix model to calculate the local light intensity along the optical axis based on the Eqs.\,\RefEqg{Bandgap44} and \RefEqg{Bandgap62} for a linear cavity for a ring cavity, respectively. Indeed, the transfer matrix formalism outlined in Secs.\,\ref{sec:TransferTransfer} and \ref{sec:TransferRing} not only allows us to calculate the bulk reflectivity of the 1D optical lattice, but also the local intensity $I\propto|\alpha_+^{(z)}+\alpha_-^{(z)}|^2$ at a point $z$ inside the lattice \cite{Slama06}.

We proceed in steps (1)~analyzing the passage of light through an optical lattice in free space, (2)~then including thermal disorder, (3)~studying the role of spontaneous emission, (4)~observing the formation of a photonic band gap, and (5)~discussing the impact of a cavity on this band structure.

\subsubsection{Light propagation through an optical lattice in free space} \label{sec:BandgapPropagationFree}

	\Fig{8.7}{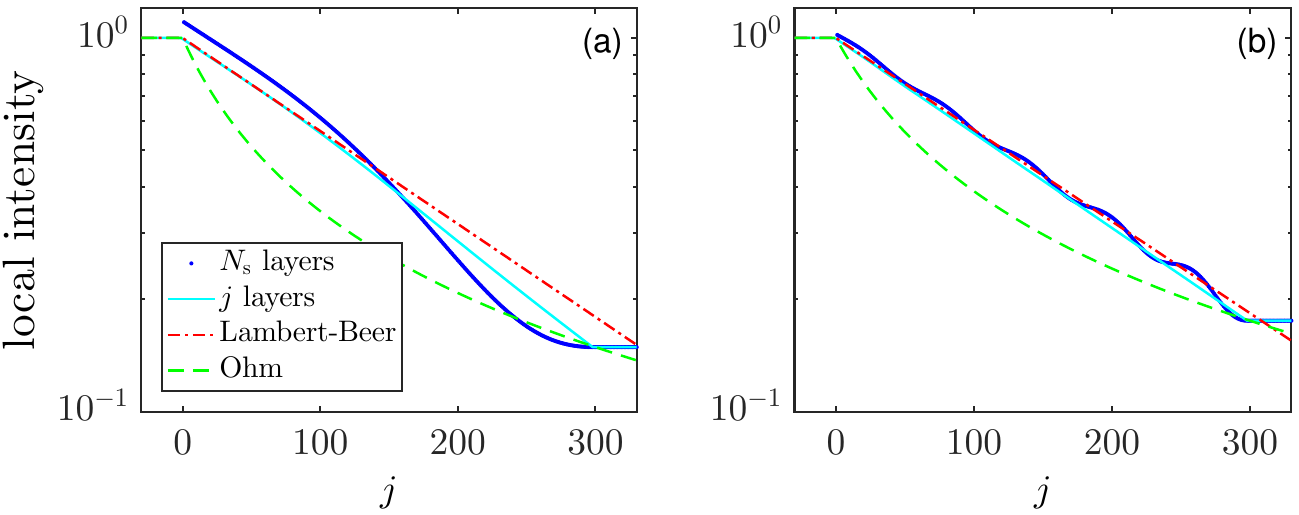}{In \emph{free space} the intensity decreases exponentially over an infinite lattice. For a finite lattice
		(here $N_\text{s}=300$), the intensity approaches a constant value at the end of the lattice. 
		The graphs show the intensity evaluated at each lattice site (solid blue lines), the exponential decay due to absorption in a 
		homogeneous cloud, i.e.~the Lambert-Beer law (dash-dotted red lines), and the transmission at the end of a lattice with $j$ layers 
		(solid cyan lines). For comparison, dashed green lines represent the hyperbolic decay following Ohm's law, 
		calibrated to the Lambert-Beer law. The parameters are $\Delta_\text{a}=\Gamma/5$, average density $n=10^{11}\U{cm^{-3}}$, 
		and for panel (a)~$\Delta_\text{lat}=2\cdot 10^8\Gamma$, while for panel (b)~$\Delta_\text{lat}=5\cdot 10^9\Gamma$.}

In free space (without cavity), the intensity drops across an optical lattice because of spontaneous emission. The thick blue curves shown in Figs.\,\RefFig{Free_Intensity} for two different lattice detunings $\Delta_\text{lat}$ represent the interference between the incident and the reflected beams, evaluated only at the exact positions of the atomic layers which are assumed to be perfectly thin. If the cloud is disordered, the intensity of a light beam crossing the cloud drops according to the Lambert-Beer law. If the cloud is ordered, the intensity shows an additional oscillatory behavior. The oscillation is due to multiple reflections between adjacent atomic layers. Indeed, at sufficiently high optical density, light can be scattered into backward direction. The interference of the light back-scattered from different atomic layers with the forward propagating light creates a modulated standing light wave whose contrast and modulation period depend on the periodicity and length of the optical lattice.

\subsubsection{Inclusion of thermal disorder} \label{sec:BandgapPropagationThermal}

Until now, we assumed in all simulations the atomic layers as perfectly thin infinite planes characterized by radially homogeneous reflectivity. In Sec.\,\ref{sec:BunchingNormalTemperature} however, we showed that at finite temperature due to thermal atomic motion the layers have a finite width given by Eq.\,\RefEqg{Bunching13}, and the 1D atomic density distribution is better described by Eq.\,\RefEqg{Bunching14}. This axial spreading and even longitudinal disorder can also be accounted for in the TMM \cite{Slama06}.

We do this by subdividing each period of the optical lattice into $N_\text{ss}=30$ sublayers of width $\Delta z$ populated with atoms according to a Gaussian distribution, i.e.~we discretize the Gaussians in the density distribution \RefEqg{Bunching14},\Eqn
	{Bandgap81}{n(z) \rightarrow \sum_{i=1}^{N_\text{ss}}n(z_i)\theta(z-z_i)\theta(z_i+\Delta z-z)~,}
where $\theta$ denotes the Heaviside function.

The quality of the atomic ordering has an important impact on absorption and phase shifts, as we will see in the following.
 	\Fig{8.7}{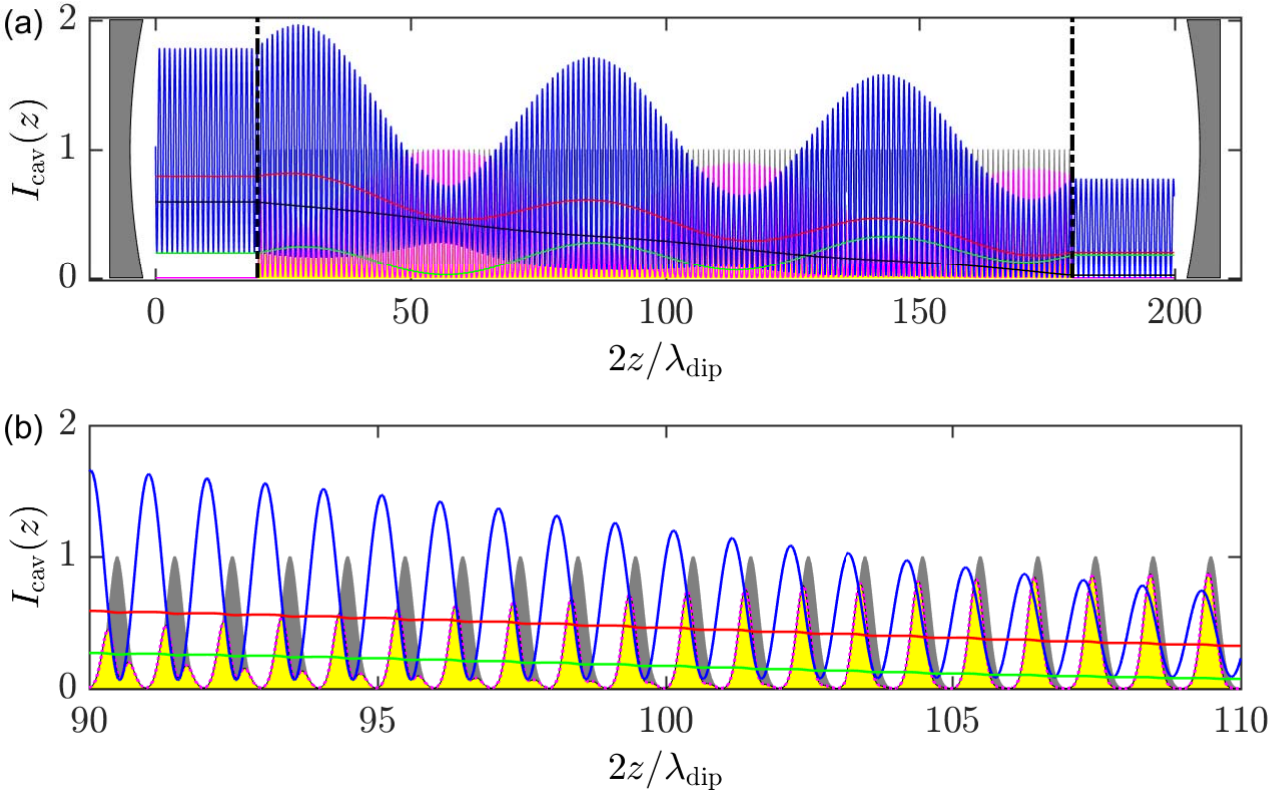}{Intensity drop along an optically dense optical lattice inside a linear cavity.
		Panel (b) shows a zoom of (a). The gray-shaded areas in (a) visualize the cavity mirrors, and the vertical black 
		dash-dotted lines delimit the optical lattice. The gray-shaded area in (b) visualizes the scaled atomic density 
		distribution assumed to be Gaussian at every lattice site. The colored lines show respectively: (red) $|\alpha_+|^2$, 
		(green) $|\alpha_-|^2$, and (blue) $|\alpha_++\alpha_-|^2$. The yellow shaded area shows $d(|\alpha_+|^2-|\alpha_-|^2)/dz$.
		The pink line shows the scaled overlap $n(z)|\alpha_++\alpha_-|^2$. The parameters differing from those given in 
		Sec.\,\ref{sec:BunchingNormalParameters} are $N_\text{s}=200$, $N_\text{ss}=30$, $\bar z=16\lambda_\text{lat}$, 
		$\Delta_\text{lat}=400\cdot 2\pi\delta_\text{fsr}$, and $\Delta_\text{c}=\Delta_\text{a}=5\Gamma$, and $R_\text{mir}=90\%$.}

\subsubsection{Spontaneous emission along the lattice} \label{sec:BandgapPropagationSpontaneous}

In a cavity, the field intensities $|\alpha_\pm^{(z)}|^2$ can vary along the optical axis due to multiple reflections between atomic layers of the optical lattice. However, because of energy conservation we would expect the total flux of energy $|\alpha_+^{(z)}|^2-|\alpha_-^{(z)}|^2$ to be the same all along the optical axis and even after transmission through the whole cavity. This is, however, not true in the presence of spontaneous emission, which is strong wherever the spatial overlap between the light intensity $|\alpha_+^{(z)}+\alpha_-^{(z)}|^2$ and the atomic density $n(z)$ is large. Hence, in case of a dense atomic cloud the intensity $|\alpha_+^{(z)}|^2-|\alpha_-^{(z)}|^2$ drops along the optical axis due to spontaneous emission,\Eqn
	{Bandgap83}{\frac{d}{dz}(|\alpha_+^{(z)}|^2-|\alpha_-^{(z)}|^2) \propto n(z)|\alpha_+^{(z)}+\alpha_-^{(z)}|^2~.}
This relationship is confirmed by the simulations exhibited in Fig.\,\RefFig{Linear_Intensity} for a linear cavity and in Fig.\,\RefFig{Ring_Intensity} for a ring cavity.

\bigskip

Similar to the case of an optical lattice in free space (cf.~Fig.\,\RefFig{Free_Intensity}), the intensity inside a cavity is not uniformly distributed along the optical axis when an optical lattice is present. However, the boundary conditions are different, because the cavity mirrors reflect a large part of the light back into the optical lattice. In the case of a linear cavity, we thus expect the formation of a standing light wave which, inside the optical lattice, is modulated due to back-scattering from the atomic layers. Additionally, the intensity drops across the optical axis due to spontaneous emission losses, as can be noticed in Figs.\,\RefFig{Linear_Intensity} and \RefFig{Ring_Intensity}.
 	\Fig{8.7}{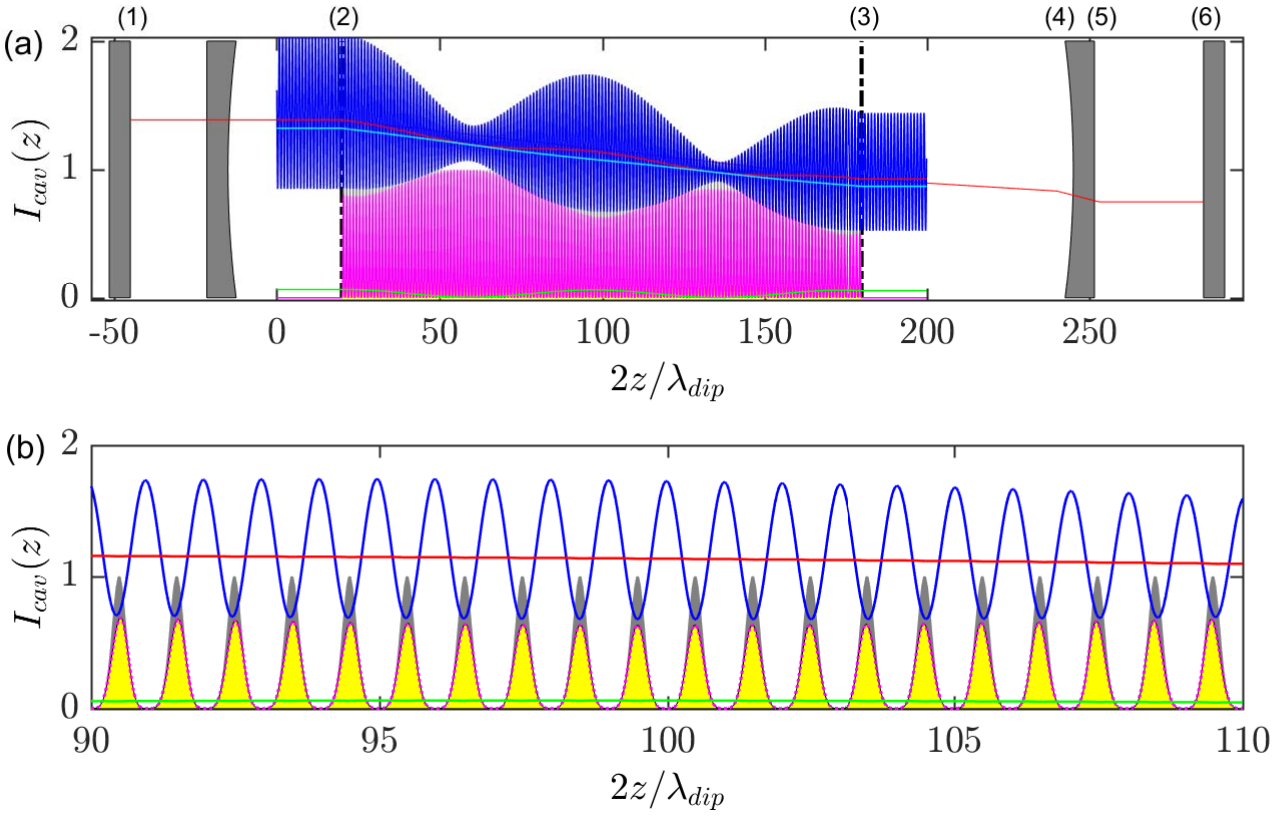}{Intensity drop along an optically dense optical lattice inside a ring cavity.
		The color coding and the parameters are the same as in Fig.~\RefFig{Linear_Intensity}.
		The numbers above the panel refer to the positions in the ring cavity indicated in Fig.\,\RefFig{Scheme2}(b).}

\subsubsection{Readjusting the cavity length} \label{sec:BandgapPropagationLock}

The power of light injected into a resonant cavity is enhanced by factor of $F/\pi$, but it is suppressed off resonance. Now, a dense optical lattice will cause the cloud's refraction index to deviate from 1 and hence the phase front of the light beam to be advanced or delayed, depending on its detuning from the atomic resonance. This modifies the resonance condition for the cavity. In practice, this can be avoided by allowing the cavity length to readjust to the resonance condition optimizing for maximum transmission, e.g.~via piezo transducers controlling the position of cavity mirrors \cite{Note01}. Note that, in the linear regime (below saturation) adjusting the resonance condition only increases the amplitude of the intracavity light, but not the profile of the intensity distribution along the optical axis.

\subsection{Cavity as a spectrum analyzer for multiple reflections} \label{sec:BandgapPhotonic}

Let us now apply the TMM to the calculation of transmission and reflection spectra, as we already did with the ODM in Sec.\,\ref{sec:BunchingNormal}. Interestingly, we find that the cavity has the tendency to fade out photonic band gaps. This can be observed in Figs.\,\RefFig{Linear_Bandgap} which show the reflection spectra of a pumped linear cavity. The spectrum in panel (a) is calculated for low cavity finesse ($R_\text{mir}\rightarrow 0$) corresponding to propagation in free space. The photonic band structure is clearly visible as a reflection band whose width depends on the lattice constant measured by $\Delta_\text{lat}$. For the panels (b-d), as the cavity finesse is gradually increased, the reflection band dissolves into a discrete spectrum of narrow resonances. Finally, for the spectrum in panel (d), the TMM and the ODM give almost the same results (not shown).

To understand this behavior, we need to remember that photonic bands result from \emph{multiple reflections} at consecutive atomic layers. Every back-reflection of the probe laser beam from a particular layer leads to a well-defined phase shift. This phase shift accumulates with the number of reflections, so that the total phase shift depends on the number of back-scattering events per round trip. At high optical density, a large number of back-scattering events are possible, so that the light reflected from the entire lattice is composed of many different phases which, in transmission, leads to destructive interference over a wide range of frequencies, thus forming a photonic stop band.
 	\Fig{8.7}{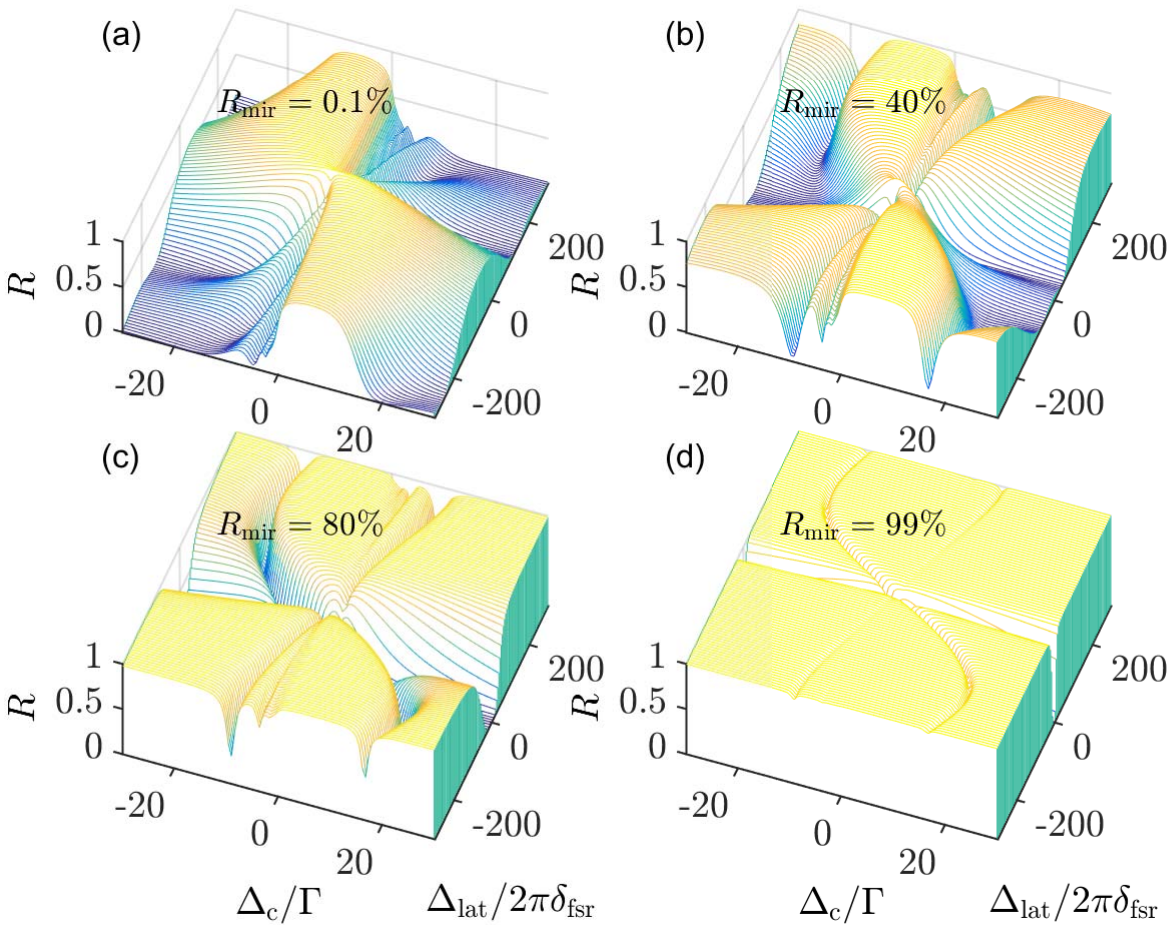}{Reflection spectra of a linear cavity showing the photonic band structure as a function of laser detuning 
		and lattice constant calculated (a)~from the ODM and (b-d)~from the TMM for various reflectivities $R_\text{mir}$ of the mirrors, 
		as indicated in the figure panels. Same parameters as in Fig.\,\RefFig{Linear_Bunching} except for $g=10\Gamma$.}

\bigskip

The presence of atoms in the cavity introduces a refraction index, which delays or advances the phase front of the light beam circulating in the cavity, which in turn modifies the resonance condition for the cavity. With the parameters specified in Sec.\,\ref{sec:BunchingNormalParameters}, the reflectivity of one atomic layer is $N_1\beta_0\approx 0.01$, corresponding to a shift on the cavity's resonance frequency of\Eqn
	{Filter02}{\delta_\text{fsr}N_1\beta_0 \approx (2\pi)\,12.7\U{MHz}~,}
which is on the order of the cavity's transmission linewidth $\kappa/2\pi\approx 3.4\U{MHz}$.

Additionally, as the total phase shift per round trip depends on the number of back-scattering events, and since the cavity can only be resonant for very specific total phase-shifts, it acts as a filter only allowing for specific numbers of reflections within the optical lattice. This number can be tuned via the cavity's length. In other words, the cavity operates like a spectrum analyzer only allowing for very specific reflection paths. For example, a cavity could be tuned to only allow for two reflections at adjacent atomic layers.

This view is supported by simulations exhibited in Figs.\,\RefFig{Linear_Filter} showing transmission spectra of a linear cavity. Panel\,(x) is obtained for vanishing finesse (no cavity). The panels (a-g) are calculated for $R_\text{mir}=80\%$ but different lengths of the cavity equally distributed between $L+\tfrac{n}{7}\tfrac{\lambda_\text{lat}}{2}$, where $n=1,2,...,7$. Finally, panel\,(y) shows the sum of all spectra exhibited in (a-g).
 	\Fig{8.7}{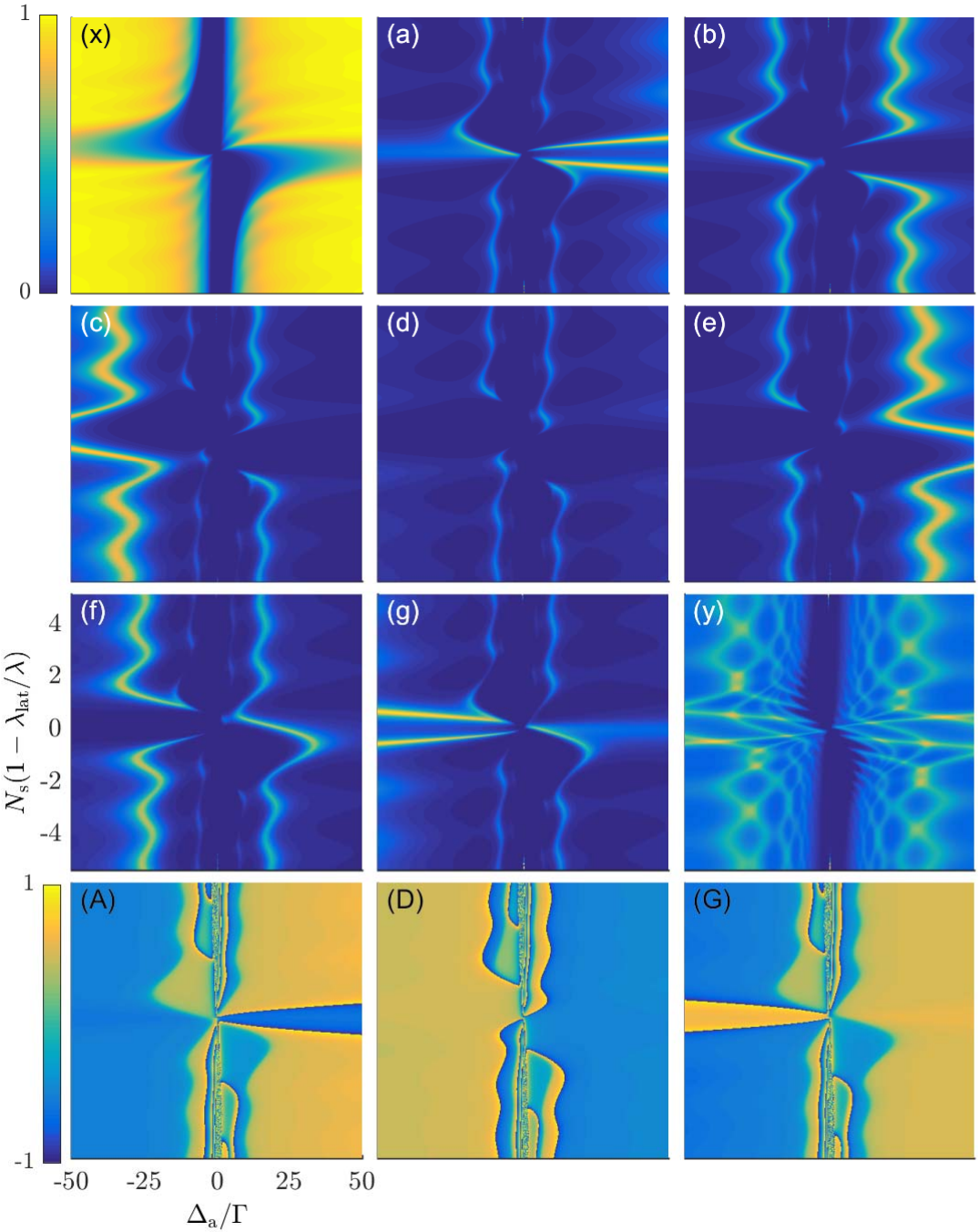}{Transmission spectra $T_+\propto|\alpha_+^\text{out}|^2$ of a linear cavity for the same parameters 
		as in Fig.\,\RefFig{Linear_Bandgap}. (x)~Photonic band structure observed for $R_\text{mir}=0.1\%$ and (a-g)~for $R_\text{mir}=80\%$. 
		The spectra in (a-g) are calculated for different lengths of the cavity. (y)~Sum of all spectra. 
		(A,D,G)~Phase profiles calculated from $\Im(\ln\alpha_+^\text{out})$ corresponding to the spectra (a,d,g).
		The abrupt transitions are caused by phase wrapping.}

The Figs.\,\RefFig{Linear_Filter} exhibit many features revealing a rich underlying dynamics: (i)~Under the influence of a linear cavity the photonic band spectrum\,(x) dissolves into narrow fringes meandering between $\Delta_\text{lat}=-\infty$ and $+\infty$ with a characteristic oscillation period corresponding to the modulation of the bunching parameter \RefEqg{Bunching12} plotted in Fig.\,\RefFig{UncommensurateBunching}(a); that is, the modulation period is given by $N_\text{s}\lambda_\text{lat}=n\lambda$ for an integer $n$. (ii)~The amplitude and width of each fringe increases with its distance from resonance at $\Delta_\text{a}=0$. (iii)~Summing up a sufficient amount of spectra, such as those of panels\,(a-g), we recover the full photonic band spectrum; that is, with an increasing amount of spectra contributing to the sum or reducing the resolution of the cavity by diminishing $R_\text{mir}$, panel\,(y) will resemble more and more panel\,(x). (iv)~Plotting the phase profiles of the transmitted light, $\Im(\ln\alpha_+^\text{out})$, in panels\,(A,D,G), we observe that the fringes separate spectral regions of opposite phases. This points to the fact that every fringe corresponds to a different number of multiple reflections: The outer fringes involve two reflections and the fringes in the inner region of the bandgap involve several.

It is interesting to note that, although across the Figs.\,(b-j) the cavity length is varied over an entire free spectral range and that for $R=80\%$ the finesse is very poor ($F=14$, $\kappa/2\pi=270\U{MHz}$), the cavity remains very selective for the numbers of reflections and basically acts like a filter for specific beam trajectories out of the multitude of possible trajectories whose interference is at the origin of the photonic band structure.

The calculations shown in Figs.\,\RefFig{Linear_Filter} were realized for symmetric coupling, as studied in Fig.\,\RefFig{Linear_Bunching}(a), in which case the normal modes are far away from the photonic band spectrum. In contrast, for antisymmetric coupling [cf.~Fig.\,\RefFig{Linear_Bunching}(b)] the normal mode splitting, vanishing close to $\Delta_\text{lat}\simeq 0$, interferes with the photonic band spectrum. Finally, we stress that similar results can be obtained for ring cavities.

\section{Conclusion \& outlook} \label{sec:Conclusion}

State-of-the-art experiments studying the interaction of optical lattices with linear or ring optical cavities are able to reach parameter regimes characterized by high optical densities \cite{Rivero23}. To describe such experiments in the limit of weak driving, we applied two theoretical approaches in this work: the Open Dicke Model (ODM) and the Transfer Matrix Model (TMM). Both models have their advantages and limitations.

\subsection{Scope of ODM and TMM} \label{sec:ConclusionODMTMM}

The ODM is perfectly suited for the description of atom-cavity interactions (e.g.~normal mode splittings) in cases of homogeneous or perfectly periodically ordered clouds. Disordering due to thermal motion or incommensurate lattice periods can be accounted for; the latter case has even practical utility for the detection of Bloch oscillations. Although this work is restricted to the low saturation regime, the ODM can also be applied in cases of high saturation \cite{Rivero23}. 

The notion of a predetermined cavity mode function shaping the atom-field coupling constant is a basic concept of the ODM: The ODM assumes that the dynamics of every atom depends on that of the other atoms only through their coupling strength to the cavity mode, but it does not depend on the relative positions of the atoms. For this reason, the ODM does not hold for the regime of \emph{high optical density}, where an atomic cloud can produce a shadow for atoms located further downstream along the path of a laser beam.

\bigskip

In contrast, in the TMM, the dynamics of an individual atomic layer (axially thin but radially extended atomic cloud) is not related to a global cavity mode function, but only to the light field amplitudes in adjacent layers. Within the linear low saturation regime and within a mean field approximation, the atomic degrees of freedom can be removed from both models, from the ODM expressed in the Eqs.\,\RefEqg{Bunching02} [respectively \RefEqg{Bunching03} for ring cavities], as well as from the TMM culminating in Eq.\,\RefEqg{Bandgap33} [respectively \RefEqg{Bandgap64}]. In this linear regime, the TMM coincides with the ODM as long as the optical density is kept low. However, the TMM also applies for high optical densities and predicts new phenomena.

Indeed, the TMM model provides insight into the formation of photonic band gaps and their spectral analysis: Light reflected from atoms in a disordered cloud suffers arbitrary phase shifts, which leads to destructive interference. In contrast, when the cloud is periodically ordered in an optical lattice and sufficiently dense, an incident light beam will suffer a discrete number of reflections and hence an integer number of possible phase shifts. In free space we observe the formation of photonic band structures. The presence of a cavity, however, imposes resonance conditions which can only be satisfied by certain paths. Tuning the cavity resonance frequency we can thus filter out specific paths and study their contribution to the formation of the photonic band.

In the bad cavity limit, the edges of photonic bands can be quite steep (scaling like $\Gamma$, while the steepness of a cavity transmission curve typically scales as $\kappa$). This could be useful for witnessing atomic ordering with much higher sensitivity than normal mode splitting.

\subsection{Future investigations} \label{sec:ConclusionFuture}

\subsubsection{Backaction of light on atomic bunching} \label{sec:ConclusionFutureNonlinear}

The presented simulations mostly assumed immobile atoms. Experimentally, this can be guaranteed by confining them in a very deep lattice. Photonic recoil induced by the probe light can then be neglected, and residual thermal motion of the ultracold cloud can be treated as done in Sec.\,\ref{sec:BunchingNormalDebunching}.

The situation is different when quantum motion is studied, for instance, the Bloch oscillations discussed in Sec.\,\ref{sec:BunchingBlochoscis}. Then, even in the tight binding limit, the optical lattice should be relatively shallow to enable quantum tunneling between adjacent lattice sites. Under such circumstances the motional dynamics of the atoms becomes very sensitive to photonic recoil. Upon (Bragg-) reflection the atomic cloud receives twice the photonic recoil, which leads to acceleration. In contrast, upon transmission of light through the atomic cloud, information on the atomic distribution is imprinted on the light beam as a phase shift, but no recoil is imparted.

\bigskip

Other interesting perspectives arise from a mutual interplay between the intracavity light field and the atomic ordering. The intracavity field amplitudes \RefEqg{Bunching07} and \RefEqg{Bunching08} depend on atomic bunching and thus are essentially governed by \emph{effective} atom numbers $N_\text{eff}\equiv Nb_{0,\pm}$. If bunching can now be made a dynamical parameter that can be controlled or manipulated by the cavity light field, then a feedback mechanism can be implemented which may me harnessed for engineering non-linear dynamics. This has been exploited in the past in a variety of systems. Prominent examples are the Collective Atomic Recoil Laser and similar systems \cite{Kruse03b,Nagorny03,Slama07} or the Dicke phase transition in which the non-linear dynamics leads to atomic self-ordering. A recent work \cite{ZhangH23} proposes to amplify Bloch oscillations of effective atom numbers via feedback from cavity fields. Control of the effective atom number can also be engineered in multimode light fields selectively interacting with several atomic ground states \cite{Suarez23,Hernandez24}.

\subsubsection{Towards a full quantum model for saturated optically dense clouds} \label{sec:ConclusionFutureGreen}

Frequency-shifts of normal mode spectra induced by saturation, as predicted by the ODM, can also be taken into account in the TMM, since with the saturation parameter given by,\Eqn
    {Conclude01}{s(\Delta_\text{a}) = \frac{2\Omega^2}{4\Delta_\text{a}^2+\Gamma^2}~,}
the polarizability can be generalized to,\Aln
    {Conclude02}{\frac{\alpha_\text{pol}}{\varepsilon_0}
        & = \frac{6\pi}{k^3}\frac{-\Gamma}{\imath+2\Delta_\text{a}/\Gamma}\frac{1}{1+s}\\
        & = \frac{6\pi}{k^3}\frac{\imath-2\Delta_\text{a}/\Gamma}
            {1+4\Delta_\text{a}^2/\Gamma^2+2\Omega^2/\Gamma^2}~.\nonumber}
It is, however, important to realize that this procedure misses saturation-induced non-linearities and bistabilities \cite{Rivero23}, which therefore are beyond the TMM. Indeed, the description of the atomic cloud as a classical medium characterized by a refractive index implies that it is not saturable. 

Consequently, effects of saturation were not treated in this work, although saturation of the atomic transition may tremendously impact the behavior of strongly pumped systems, in particular, when a narrow atomic transition is used. As long as the optical density stays low, the non-linear ODM can be used \cite{Rivero23}, but for high optical densities this model fails, as shown in this work, and the same is true for generalizations based on input-output theories \cite{Gardiner85} not containing atomic variables. 

On the other hand, the problem with the TMM is that, when the atomic medium becomes transparent under saturation, it reduces its optical density, which breaks the linearity of the transfer matrix concatenation procedure. Below saturation one can define a generalized mode function shaped not only by the cavity geometry but also by the positions and reflectivities of intracavity scatterers. This is what the TMM provides: a bulk scattering matrix, which only depends on experimental parameters and whose response function delivers the dynamics of the system, even if the atomic medium is dense. Above saturation this is not possible, because transmission and reflection dynamically depend on the states of the atoms, and therefore the TMM must fail to describe saturation and quantum correlations.

\bigskip

In any case, one must resort to more sophisticated models handling quantum mechanically the interaction between a dense cloud and a saturating beam inside a cavity with partially reflecting mirrors. The ODM only allows the atoms to interact via the cavity field. This however is not a good assumption in the dense regime, which is characterized by the fact that atoms can absorb radiation emitted by other atoms, a fact that is accounted for by the TMM, where subsequent atomic layers can exchange photons directly via reflections. In free space these direct interactions are included as dipole-dipole interactions via \cite{Lehmberg70},\Eqn
	{Conclude03}{\hat H_\text{Ising} = \sum_{i\neq j}\Delta_{ij}\hat\sigma_j^+\hat\sigma_i^-~,}
where $\hat\sigma_j^\pm$ denote the standard Pauli matrices decribing (de-)excitation of the $j^\text{th}$ atom and $\Delta_{ij}$ the interatomic coupling strengths. 

Clearly, the inclusion of dipole-dipole interactions is beyond the Dicke model. Additionally, in a cooperative environment, such as a cavity, these interactions are modified, as well as the interaction of the atoms with the vacuum modes. The starting point for setting up the Hamiltonian must, therefore, be \emph{before} the Weisskopf-Wigner treatment of spontaneous emission leading to the Lindbladian \cite{Milonni-94}. Introducing the field operators $\hat a_{\k\lambda}$ and $\hat a_{\k\lambda}^\dagger$ for the creation and annihilation of photons in a radiation mode ($\k,\lambda$), the collective interaction Hamiltonian for an ensemble of atoms interacting via dipole-dipole interactions, driven by laser beam and coupled to every vacuum mode is,\Eqn
	{Conclude04}{\hat H = \hbar\sum_{\k,\lambda}\sum_j(\hat\sigma_j^++\hat\sigma_j^-)
		(g_{\k\lambda}\hat a_{\k\lambda}+g_{\k\lambda}^\ast\hat a_{\k\lambda}^\dagger)~,}
where the coupling strength $g_{\k,\lambda}$ is shaped by the presence of a cooperative environment. That is, the boundary conditions imposed by the cavity mirrors make that $g_{\k\lambda}$ be anisotropic \cite{Heinzen87,Heinzen87c} and spectrally modulated by the cavity's resonance conditions. A proper Weisskopf-Wigner treatment then allows to calculate the density of states shaped by the presence of the cavity, the cavity's Purcell factor (also called cooperativity or structure factor), and the cooperative Lamb shift. This however is beyond the scope of the present paper.

Despite its difficulties, a full quantum model working for dense ensembles of saturable scatterers inside an optical cavity is highly desirable. To give just one example within the scope of this work, it is conceivable that in the non-linear regime the cavity-filtering of photons having suffered exactly two reflections at adjacent atomic layers can be harnessed for protocols for the generation of quantum correlations leading to superradiant lasing \cite{Meiser09,MaierTh14,Norcia16c,Schaeffer20}.


\end{document}